\documentclass[a4paper, 11pt]{article}
\pdfoutput=1

\usepackage{jcappub}
\usepackage{graphicx}
\usepackage{booktabs}

\newcommand{\keVee}{\mathrm{keV}_{\mathrm{ee}}}

\title{
Quantifying the evidence for dark matter in CoGeNT data
}

\author[a]{Jonathan H.~Davis,}
\author[a]{Christopher McCabe,}
\author[a,b]{and C\'eline B\oe hm}

\affiliation[a]{Institute for Particle Physics Phenomenology, Durham University, South Road, Durham, DH1 3LE, United Kingdom}
\affiliation[b]{LAPTH, U. de Savoie, CNRS,  BP 110, 74941 Annecy-Le-Vieux, France}

 \emailAdd{j.h.davis@durham.ac.uk}
 \emailAdd{christopher.mccabe@durham.ac.uk}
  \emailAdd{c.m.boehm@durham.ac.uk}

\abstract{
We perform an independent analysis of data from the CoGeNT direct detection experiment to quantify the evidence for dark matter recoils. We critically re-examine the assumptions that enter the analysis, focusing specifically on the separation of bulk and surface events, the latter of which constitute a large background. This separation is performed using the event rise-time, with the surface events being slower on average. We fit the rise-time distributions for the bulk and surface events with a log-normal and Pareto distribution (which gives a better fit to the tail in the bulk population at high rise-times) and account for the energy-dependence of the bulk fraction using a cubic spline. Using Bayesian and frequentist techniques and additionally investigating the effect of varying the rise-time cut, the bulk background spectrum and bin-sizes, we conclude that the CoGeNT data show a preference for light dark matter recoils at less than $1\sigma$.
}

\arxivnumber{IPPP/14/32, DCPT/14/64}

\begin{document}
\maketitle
\flushbottom

\section{Introduction}

In many theories of particle dark matter (DM), the DM particle has a small interaction with the Standard Model particles. Direct detection experiments are searching for evidence of this interaction.  Of particular interest in the past few years have been the results from the  DAMA/LIBRA~\cite{Bernabei:2013xsa}, CoGeNT~\cite{Aalseth:2010vx,Aalseth:2011wp,Aalseth:2012if,Aalseth:2014eft}, CRESST-II~\cite{Angloher:2011uu,Brown:2011dp} and CDMS-Si~\cite{Agnese:2013rvf} experiments, which have observed an excess number of events above naive background expectations. These experiments have been interpreted as providing evidence for DM elastically scattering on nucleons with a mass around $10\,$GeV. However, this interpretation is not supported by the results of CDEX~\cite{Yue:2014qdu}, CDMS-II~\cite{Ahmed:2010wy}, EDELWEISS-II~\cite{Armengaud:2012pfa}, LUX~\cite{Akerib:2013tjd}, SuperCDMS~\cite{Agnese:2014aze}, XENON10~\cite{Angle:2011th,Frandsen:2013cna} and XENON100~\cite{Aprile:2012_225}, for example, thus calling for a thorough investigation of available data.

The CoGeNT collaboration have publicly released their 1129~live-days dataset, which contains events spanning the period between  4th December 2009 and 23rd April 2013~\cite{Aalseth:2014eft}. This dataset shows an exponential rise in the number of events at energies below~$1~\keVee$, where a low-mass DM recoil signal is expected to occur. 
The CoGeNT dataset also shows $\sim 2.2 \sigma$ evidence for the expected $\sim10\%$ annual modulation in the event rate induced by DM recoils~\cite{Drukier:1986tm,Freese:1987wu}. This may also be consistent with a DM recoil signal, but appears to require significant departures from a Maxwellian velocity distribution~\cite{Aalseth:2014eft, Fox:2011px,McCabe:2011sr,Frandsen:2011gi,Arina:2011zh,HerreroGarcia:2011aa}. In light of this, in this paper we focus only on the unmodulated data, which exhibits the low-energy exponential rise, and leave the analysis of the much smaller annual modulation signal to future work. Our aim here is to quantify the evidence for a DM signal in the unmodulated dataset.

A limitation of the CoGeNT experiment is that the data contains both bulk and surface events; the surface events constitute a background that mimics a DM recoil signal at low energy so should be removed before searching for a DM signal. In section~\ref{sec:risetimes} we discuss how the bulk and surface events can be distinguished by their `rise-time' since surface events typically have a longer rise-time than bulk events. Unfortunately, the populations of bulk and surface events overlap significantly at low-energy because the difference between the bulk and surface events rise-time is small. This means that in the low-energy region of interest for DM searches, only a fraction of the observed events are bulk events; the fraction of bulk events (the `bulk fraction') must be determined before the evidence for DM can be quantified. 

In section~\ref{sec:1129} we characterise the bulk fraction as a function of energy using cubic splines (owing to the absence of a theoretically motivated function). This allows us to robustly incorporate the bulk fraction uncertainties in our Bayesian and frequentist analyses. Following previous analyses~\cite{Aalseth:2012if,Aalseth:2014eft,Aalseth:2014jpa} we first model the bulk event population with a log-normal distribution. We then investigate fits with a Pareto distribution that gives a better fit to pulser data (which mimic bulk events) at larger values of the rise-time. We also test the robustness of our conclusions by considering variations in the DM search region (by investigating alternative rise-time cuts), the bulk background spectrum and bin-sizes.

As our conclusions differ from the results of the CoGeNT analysis in ref.~\cite{Aalseth:2012if}, in section~\ref{sec:807} we reanalyse the older 807 live-days dataset used in that analysis. We summarise all of our findings in section~\ref{sec:sum} and conclude in section~\ref{sec:conc}. A number of appendices give cross-checks of our results and technical details related to our analysis.

\section{Surface and bulk events \label{sec:risetimes}}

The CoGeNT experiment employs a detector based on a p-type germanium semiconductor. The bulk of the detector is a pure p-type semiconductor while the surface is an inert `dead layer' (this is the location of the electrical contact for the detector). Between the dead layer and the bulk is a $\sim 1$~mm thick `transition layer'. As the transition region is on the outside of the detector module, we follow the CoGeNT collaboration and define events occurring in the transition region as `surface events'.

In this transition region, the charge collection efficiency is less than one. Because of this, the reconstructed energy of surface events will be smaller than the true energy of the incoming particle. Specifically, this implies that high energy gamma-rays (or other backgrounds) which scatter in the transition region will be reconstructed as low energy events, resulting in a build-up of low-energy events~\cite{Aalseth:2014jpa}. This is a problem because the signal from elastically scattering DM particles also has a characteristic rise at low energies so surface events can mimic a DM signal.

The transition region is only $\sim 1$~mm thick compared to the cm scale bulk region so the volume of the transition region is much smaller than that of the bulk. Since DM particles are weakly-interacting, they are significantly more likely to scatter in the larger bulk region. Hence, the removal of the surface events will reduce the number of background events that mimic a DM signal while leaving the DM rate essentially unaffected.

\begin{figure}[t!]
\centering
 \includegraphics[width=0.65\textwidth]{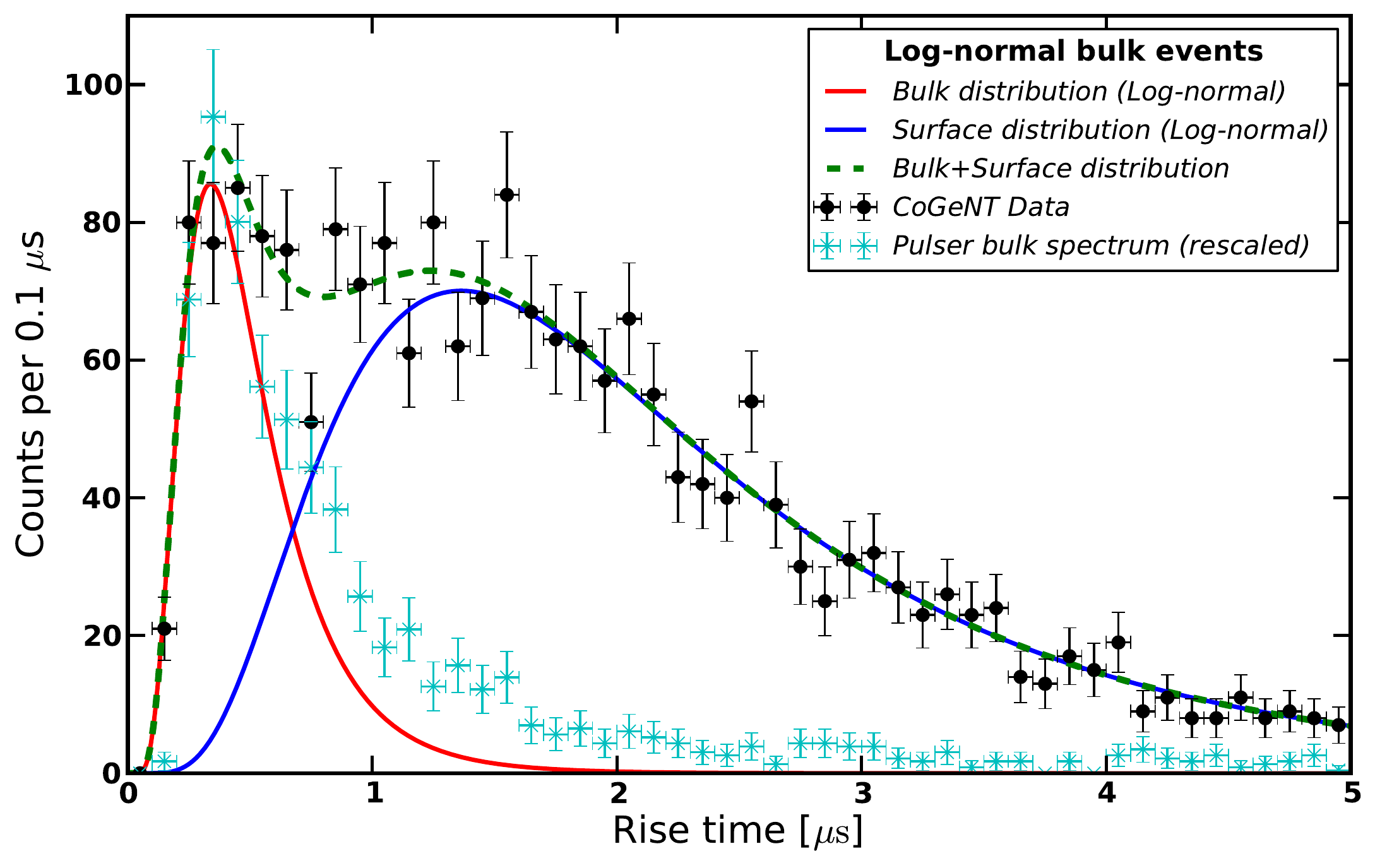}
  \includegraphics[width=0.65\textwidth]{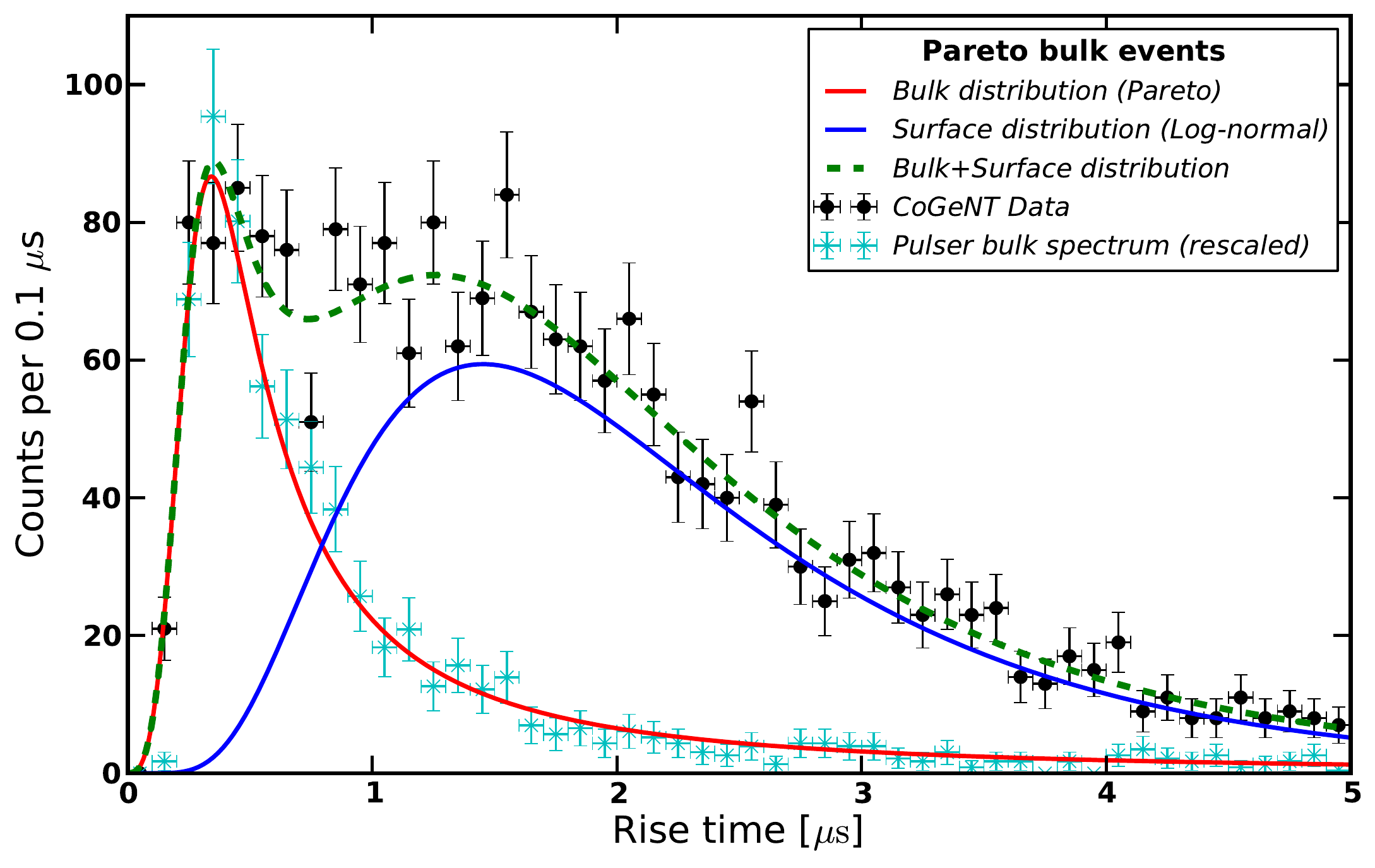}
\caption{Bulk (red solid line) and surface (blue solid line) event distributions as a function of rise-time for energies between $0.5$ keV$_{\mathrm{ee}}$ and $0.9$ keV$_{\mathrm{ee}}$. The pulser data (cyan points) mimics the behaviour of bulk events. In the upper panel both populations are fit with log-normal distributions, while in the lower panel the bulk population is fit with a Pareto distribution. Both the log-normal and Pareto bulk distributions (when summed with the surface distribution) give good fits to the CoGeNT data (black points) but the slower fall off of the Pareto distribution at large rise-time is better able to fit the pulser data.}
\label{fig:lowE_risetimes}
\end{figure}

For these reasons, to search for a DM signal it is imperative that the surface event population is separated and removed from the bulk event population. This separation is possible because surface events are characterised by a longer duration (the `rise-time') than those in the bulk. Previous analyses~\cite{Aalseth:2014eft,Aalseth:2014jpa,Aalseth:2012if} assumed that the bulk and surface event populations both follow a log-normal distribution. This assumption is motivated by the fact that the build up of charge carriers in the detector after an energy deposit is geometric i.e.~more charge carriers implies a faster rate of increase. The overlap of the two population is energy dependent and is clearer at higher energies (see~\cite{Aalseth:2012if,Aalseth:2014jpa}). The black data points in figure~\ref{fig:lowE_risetimes} show the rise-time distribution of events in the low-energy range between $0.5~\keVee$ and $0.9~\keVee$ (the range of interest for DM searches). Here the separation between bulk and surface events is not as clear. The upper panel of figure~\ref{fig:lowE_risetimes} shows that the sum of the log-normal distributions from the bulk events (red solid line) and the surface events (blue solid line) give a good fit to the CoGeNT data points.

Also shown in figure~\ref{fig:lowE_risetimes} is the rise-time distribution from fast electronic pulser events, which mimic the behaviour of bulk events~\cite{Aalseth:2014eft2}. The electronic pulser data is shown by the cyan data points (after a suitable renormalisation). The upper panel of figure~\ref{fig:lowE_risetimes} shows that these events are poorly fitted by the log-normal distribution because of the large tail of events at high rise-times. This motivates us to consider other models of the bulk event distribution. Specifically, we consider a generalised Pareto distribution whose PDF we give in appendix~\ref{app:dists}. This distribution has a parameter that allows a better fit to the tail of the pulser data, as shown in the lower panel of figure~\ref{fig:lowE_risetimes}.

Both the log-normal and Pareto fits for the bulk events (when summed with the log-normal fit to the surface events) give a comparable quality of fit to the CoGeNT data. For example, the fit with the bulk events modelled with a Pareto distribution gives a $\chi^2 = 37.3$, while with the log-normal gives $\chi^2 = 36.6$.

In light of this, in the remainder of our analysis, we consider both the log-normal and Pareto distribution for the bulk rise-time distribution. We always fit the surface events with a log-normal distribution, as we are primarily concerned with the overlap of the bulk and surface event distributions; this occurs at smaller rise-times when the difference between the log-normal and Pareto distributions is small. Finally, the pulser data is used only as a motivation to consider the Pareto distribution: we do not fit to it in our analysis, fitting only to the CoGeNT data itself.

\section{Analysis of the full 1129 live-days dataset}
\label{sec:1129}

Having demonstrated how to separate bulk events from surface events (which are dominantly background events that rise at low-energy), in this section we discuss our analysis of the full 1129 live-days CoGeNT dataset. Initially, we describe our procedure to remove the surface events leaving the bulk-only event spectrum. Performing a fit to this spectrum will allow us to quantify the evidence for DM in the CoGeNT data. We begin by assuming that the bulk population is modelled by a log-normal distribution and take an energy-independent rise-time cut \mbox{$\tau_{\mathrm{cut}} = 5 \, \mu \mathrm{s}$}. This value is conservative as it captures the majority of the total number of events in the $0.5$ to~$0.9$~keV$_{\mathrm{ee}}$ energy bin, where the DM signal is maximal (cf.~figure~\ref{fig:lowE_risetimes}). Our analysis procedure with these assumptions is presented in sections~\ref{sec:rt_func} to~\ref{sec:mar}. We then consider variations in these assumptions to test the robustness of our results: in section~\ref{sec:pareto} we model the bulk distribution with a Pareto distribution; in sections~\ref{sec:cuts} and~\ref{sec:cuts_Edep} we take an energy-independent rise-time cut at $\tau_{\mathrm{cut}}=0.4\, \mu \mathrm{s}$ and an energy-dependent rise-time cut respectively.

\subsection{Fitting the bulk fraction}
\label{sec:rt_func}

As figure~\ref{fig:lowE_risetimes} demonstrates, in a given energy range only a fraction of the events below the rise-time cut $\tau_{\rm{cut}}=5 \, \mu \mathrm{s}$ are bulk events. We refer to this fraction of events as the `bulk fraction'~$\mathcal{R}$. To remove the surface events contamination from the measured spectrum, we need to determine the bulk fraction as a function of energy. To do this, we first determine $\mathcal{R}_j$ in equally spaced energy bins $E_j$ between $0.5$ and $2.9$~keV$_{\mathrm{ee}}$ and then fit a function $\mathcal{R}(E)$ to these values. In each bin~$j$, we find the best-fit log-normal distributions for the surface and bulk populations (as in section \ref{sec:risetimes}) and integrate up to the rise-time cut~$\tau_{\mathrm{cut}}$ to calculate the bulk fraction
\begin{equation}
\mathcal{R}_{j} = \left[\frac{\text{Number of bulk events} < \tau_{\mathrm{cut}}}{\text{Total number of events} < \tau_{\mathrm{cut}}}\right]_{j}.
\label{eqn:Rbin}
\end{equation}

\begin{figure}[t!]
\centering
\includegraphics[width=0.65\textwidth]{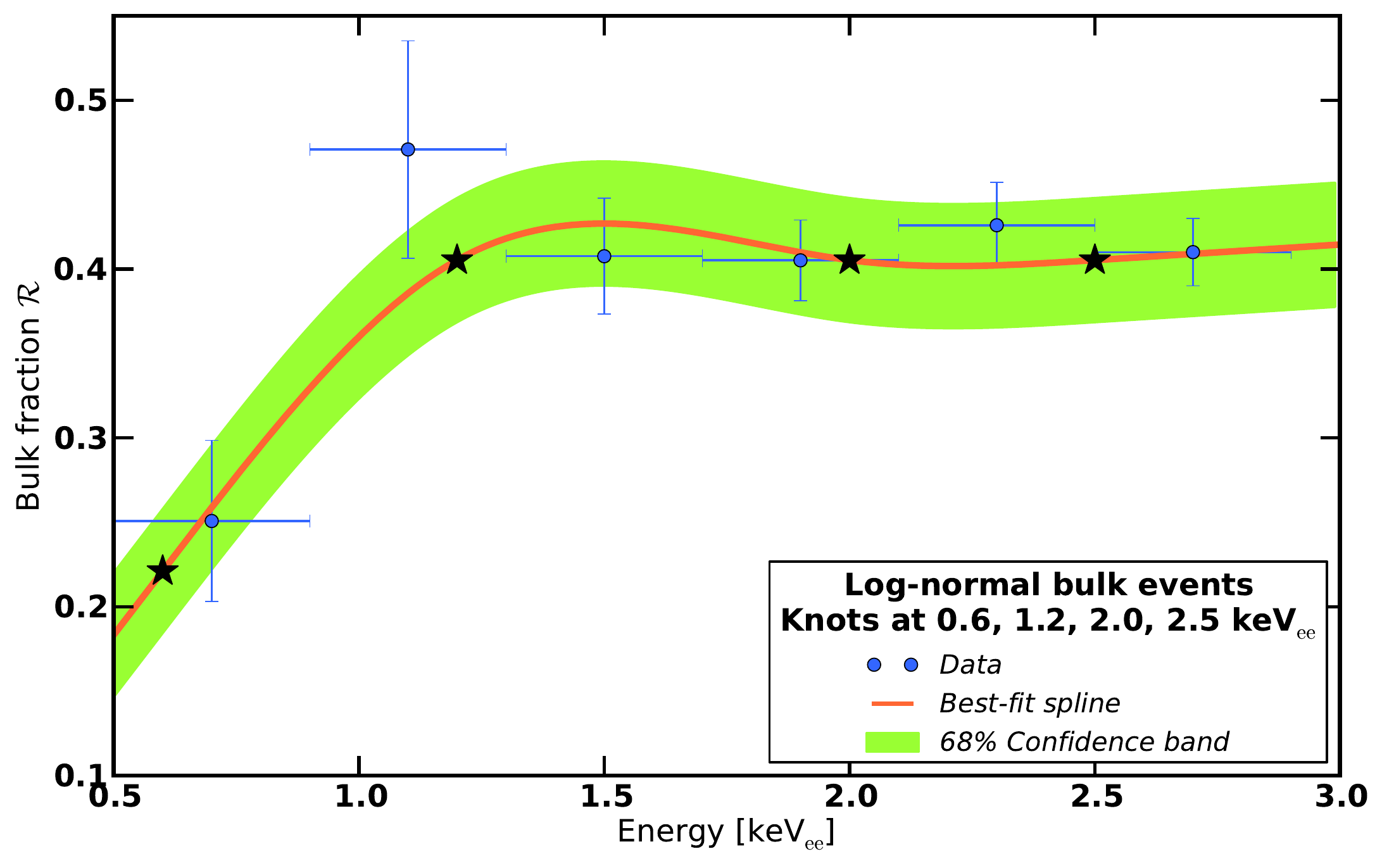}
\caption{The fraction of bulk events (`bulk fraction') obtained from log-normal fits to the rise-time data. The red solid line is the best-fit cubic spline for the knot positions indicated by the black stars. The green band represents the $68\%$ uncertainty. At low energies only around $20 \%$ of the events are from the bulk of the detector. This rises to $\sim 40 \%$ at higher energies.}
\label{fig:bulk_frac}
\end{figure}

The value of $\mathcal{R}_{j}$ and its $1\sigma$ uncertainty $\sigma_j$ for each bin are shown in figure~\ref{fig:bulk_frac}. We have used a bin width of 0.4~keV$_{\mathrm{ee}}$ to obtain six values of $\mathcal{R}_{j}$. The error-bars on $\mathcal{R}_{ j}$ arise from the uncertainty in the six parameters used to fit the two log-normal distributions. In appendix~\ref{app:bin_size} we have checked that our choice of~0.4~keV$_{\mathrm{ee}}$ for the bin width does not bias our results.

Given that there is no theoretically motivated function for $\mathcal{R}(E)$, we determine it with a cubic spline fit. This is similar to the method used by the XENON100 collaboration to parameterise the functional form of the relative scintillation efficiency L$_{\mathrm{eff}}$~\cite{Davis:2012vy,Aprile:2012_225,Manalaysay:2010mb}. The result of our cubic spline fit is shown in figure~\ref{fig:bulk_frac}. The red solid curve indicates  the best-fit cubic spline along with the associated $1\sigma$ uncertainty band (containing all splines within $1\sigma$ of the best-fit) for the choice of knots shown by the black stars. We have performed this fit using a Gaussian likelihood function
\begin{equation}
\mathcal{P}(d_{\mathcal{R}} | \mathcal{R}(E)) = \exp \left[ - \sum_j \frac{(\mathcal{R}_j - \mathcal{R}(E_j))^2}{\sigma_j^2} \right],
\label{eqn:risetime_like}
\end{equation}
where $d_{\mathcal{R}}$ refers to the bulk fraction data, $\mathcal{R}_j$ is the bulk fraction and its uncertainty $\sigma_j$ for each energy bin $E_j$ and $\mathcal{R}(E_j)$ is the value of the spline at that point. The best-fit spline is the one which maximises this likelihood. The more a spline $\mathcal{R}(E)$ deviates from the data, the smaller its likelihood will be.

The cubic spline fit shown in figure~\ref{fig:bulk_frac} is for fixed knot positions. In the remainder of our analysis we also allow the energy of the two lowest-energy knots to vary. The numerical implementation of this is discussed in appendix~\ref{sec:mar_details}. The energy position of the other knots makes little difference to our analysis as they are beyond the energy region of interest for DM recoils. Therefore we do not consider variations in their position.

\begin{figure}[t]
\centering
\includegraphics[width=0.65\textwidth]{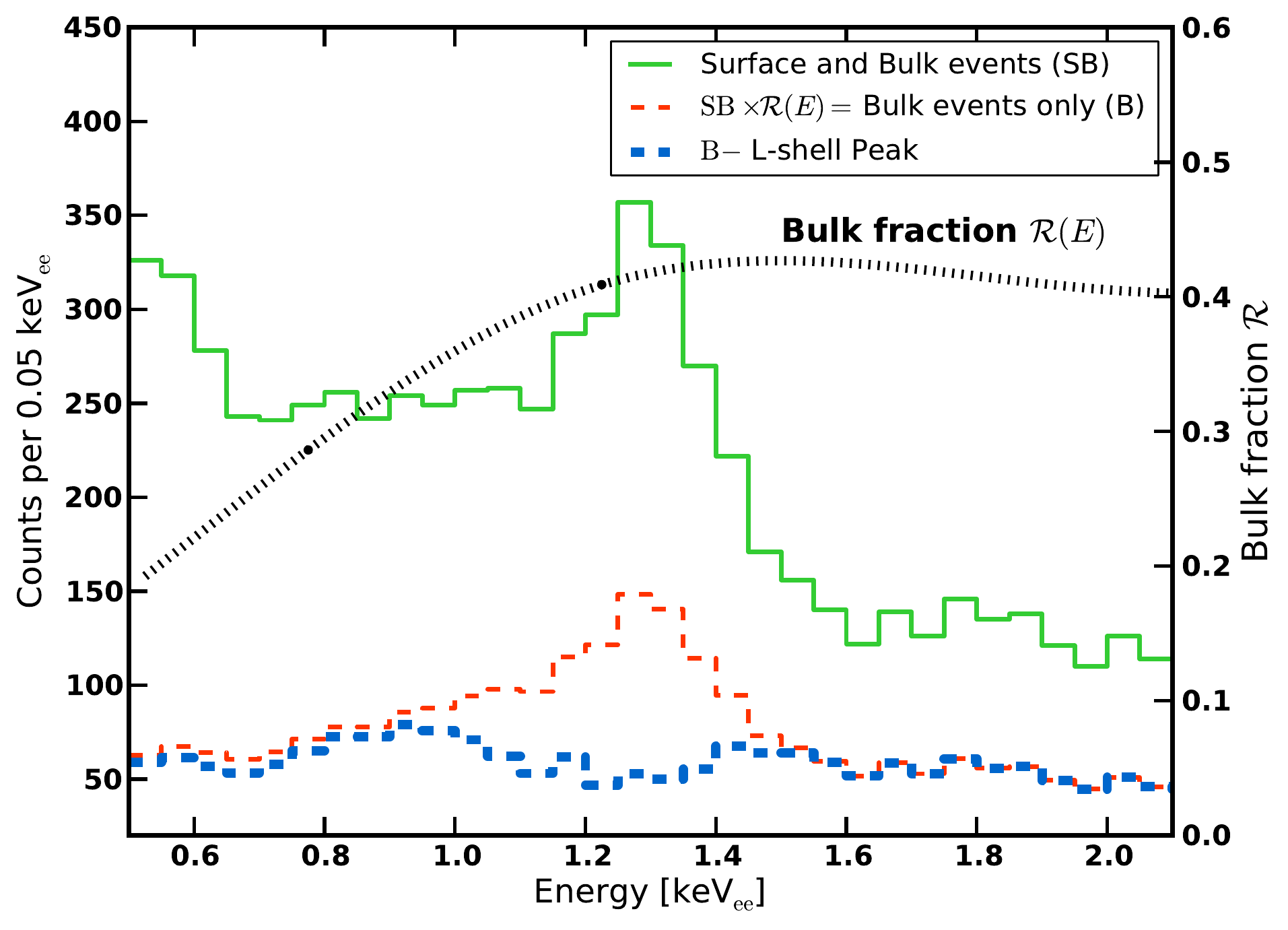}
\caption{A graphical demonstration of the procedure to remove the surface events. Upon multiplying the total spectrum (green solid) by the bulk fraction $\mathcal{R}(E)$, we are left with the red dashed spectrum, which contains only bulk events. After removing the L-shell peak at $\sim1.3~\keVee$, the resulting blue dashed spectrum is used to search for a DM signal.}
\label{fig:cogent_surf_removal}
\end{figure}

\subsection{Removing the surface events}
\label{sec:RE}

We now use the function $\mathcal{R}(E)$ to remove the surface event contamination, leaving a spectrum of purely bulk events. This procedure is shown in figure~\ref{fig:cogent_surf_removal} using the best-fit spline from figure~\ref{fig:bulk_frac} for $\mathcal{R}(E)$. The spectrum of surface and bulk events (green solid line) is multiplied by the bulk-fraction $\mathcal{R}(E)$ to obtain the bulk-only spectrum (red dashed line). We then remove the L-shell peak at $E\sim1.3~\keVee$, as discussed in appendix \ref{sec:bg}, to leave the blue dotted spectrum. This is the spectrum which can be analysed for a DM signal.

However, to robustly quantify the evidence for a DM signal, we should not only consider the best-fit spline but also all the other splines within e.g.~the 1$\sigma$ or 2$\sigma$ uncertainty bands. Since each spline implies a different bulk fraction, the resulting bulk-only spectrum associated with all these splines are different from that obtained using the best-fit spline. Hence, the evidence for a DM recoil signal depends strongly the choice of spline fit for $\mathcal{R}(E)$ (especially at energies around 0.5 keV$_{\mathrm{ee}}$). We therefore need to marginalise over all splines to obtain a robust unbiased result.

\subsection{Marginalising over the bulk fraction}
\label{sec:mar}

A robust (unbiased) analysis should be independent of the choice of spline for $\mathcal{R}(E)$.
Therefore, we consider all possible $\mathcal{R}(E)$ splines by marginalising over these different choices. We start by defining the posterior distribution $\mathcal{P}(m,\sigma | d)$, which tells us how much the data prefer a given DM recoil signal over background for all choices of $\mathcal{R}(E)$. This is defined using Bayes' theorem,
\begin{align}
\mathcal{P}(m,\sigma | d) \mathcal{P}(d) &= \mathcal{P}(d | m,\sigma) \mathcal{P}(m,\sigma)\\
&= \int \mathcal{P} (d_E | m,\sigma,\mathcal{R}(E)) \mathcal{P}(\mathcal{R}(E)) \mathcal{P}(m,\sigma) \, \mathrm{d} \mathcal{R}\,,
\label{eqn:mar1}
\end{align}
where $d=d_{E}+d_{\mathcal{R}}$ refers to the sum of the data from the bulk energy spectrum and bulk fraction respectively, $\mathcal{P}(m,\sigma) \equiv \mathcal{P}(\sigma) \mathcal{P}(m)$ is the prior for the theoretical parameters $m$ and $\sigma$, the mass and cross-section of the DM particle, and $\mathcal{P}(\mathcal{R}(E))$ is the prior for $\mathcal{R}(E)$.  The form of the prior $\mathcal{P}(\mathcal{R}(E))$ is determined from the likelihood from the spline fits of section~\ref{sec:rt_func}.

\begin{figure}[t]
\centering
\includegraphics[width=0.99\textwidth]{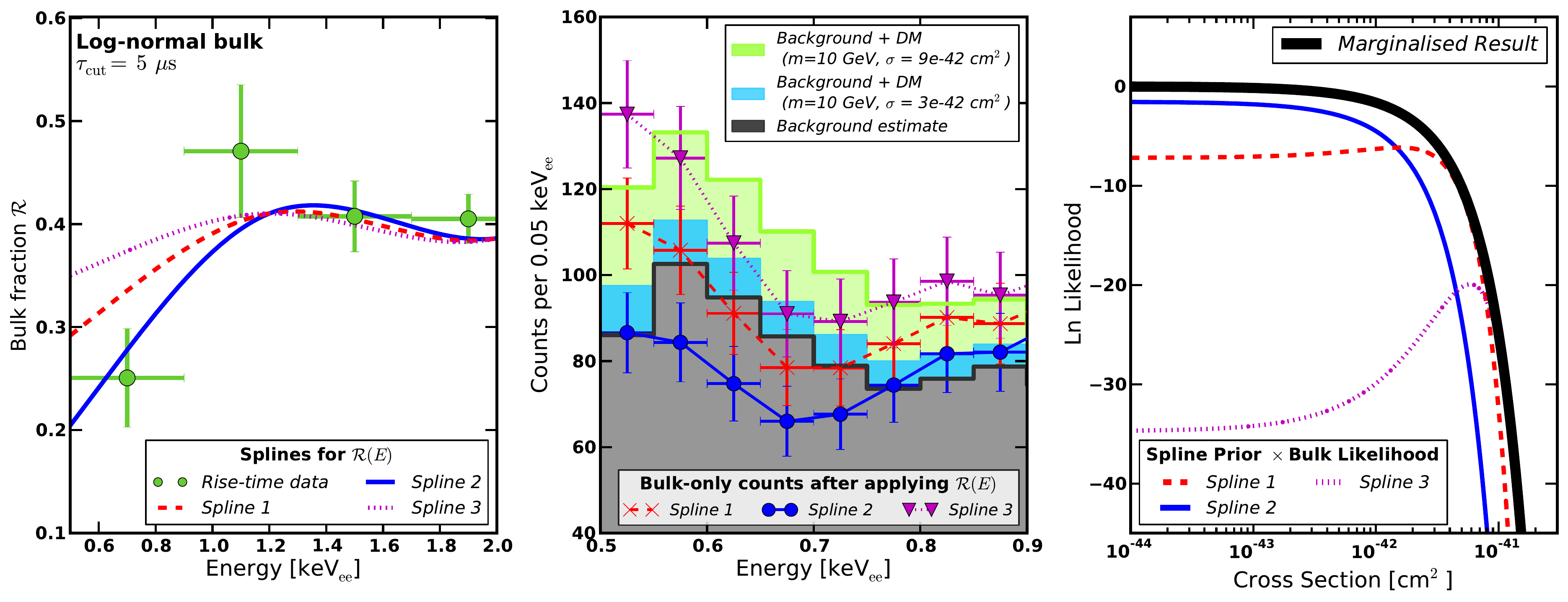}
\caption{
Analysis procedure for a 10 GeV DM particle when the bulk events are modelled with a log-normal distribution and $\tau_{\rm{cut}}=5~\mu \mathrm{s}$. The left panel shows three spline choices for the bulk fraction~$\mathcal{R}(E)$. The central panel shows the resulting bulk-only data for each spline together with CoGeNT's background estimate (black) and the number of events expected from signal+background (green and blue). Spline 1 and spline 3 result in an excess above the background estimate at low energy. The right panel shows the likelihoods from the fits in the left and central panels. The peaks for spline 1 and spline 3 indicate a preference for non-zero values of $\sigma$ with different significances. Spine 2 results in no excess (central panel) and no preference for non-zero $\sigma$ (right panel). The solid black line in the right panel is the marginalised posterior, obtained by marginalising over all spline choices. It does not posses a maximum value, indicating a vanishing preference for a DM recoil signal over the background estimate.}
\label{fig:mar_fig}
\end{figure}

The function $ \mathcal{P} (d_E | m,\sigma,\mathcal{R} )$ is the likelihood for the bulk-only data and it tells us how much the data prefer a DM recoil signal compared to the background-only scenario. If the data prefer a DM+background fit over the background-only scenario, then this should result in a peak in this likelihood at the preferred value of the cross-section $\sigma$. 
This likelihood takes the form of a Poisson function,
\begin{equation}
 \mathcal{P} (d_E | m,\sigma,\mathcal{R}(E) ) = \prod_{i=1}^N \frac{\lambda_i^{n_i} e^{-\lambda_i}}{n_i!},
\label{eqn:poisson_l}
\end{equation}
where $i$ runs over the $N$ energy bins (of width $0.05$~keV$_{\mathrm{ee}}$), $n_i$ is the number of events in each bin and $\lambda_i = f_i(m,\sigma) + b_i$ is the sum of signal $f(m,\sigma)$ and background $b$ in each bin. For the signal, we use the standard DM elastic recoil spectrum (described in appendix~\ref{sec:recoil}) and the background refers to that from bulk events only; we assume that all surface events are removed by the bulk fraction function~$\mathcal{R}(E)$. In our analysis we use the spectral shape of CoGeNT's background estimate given in~\cite{Aalseth:2012if}, corrected for the increased exposure. Appendix~\ref{sec:bg} shows that we obtain similar results with an alternate background estimate from~\cite{Aalseth:2014jpa} (cf.~figures~\ref{fig:mar_fig} and~\ref{fig:likes_flatbg}). Our likelihood analysis is performed over the energy range 0.5 to $3.0~\keVee$, the same range we considered when determining $\mathcal{R}(E)$ (cf. figure~\ref{fig:bulk_frac}). We investigate variations in this energy range in appendix~\ref{sec:energy_range} and find that the results are robust against these variations.

For practical purposes, our marginalisation reduces to the following discrete sum,
\begin{equation}
\mathcal{P}(m,\sigma | d) \propto \sum_{\mathrm{splines}} \mathcal{P} (d_E | m,\sigma,\mathcal{R}(E))  \mathcal{P}(d_{\mathcal{R}} | \mathcal{R}(E)) \mathcal{P}(m,\sigma) ,
\label{eqn:mar_discrete}
\end{equation}
where we sum over all splines. Appendix~\ref{sec:mar_details} describes in detail our numerical implementation of this procedure. For a particular choice of spline, $\mathcal{P} (d_E | m,\sigma,\mathcal{R}(E))$ gives the quality of the DM recoil signal fit to the bulk energy spectrum. This is weighted by $\mathcal{P}(d_{\mathcal{R}} | \mathcal{R}(E))$, which tells us how well this spline fits the bulk fraction $\mathcal{R}$ from our rise-time fits.  We assume that $\mathcal{P}(m,\sigma)$ is flat so that all values of $m$ and $\sigma$ are considered equally.

The process used to calculate $\mathcal{P}(m,\sigma | d)$ for $m = 10$~GeV is shown graphically in figure~\ref{fig:mar_fig} for three individual splines choices. Each spline is the result of a fit to the bulk fraction shown in the left panel and leads to a different spectrum of bulk only events, shown in the central panel.  One immediate observation is that choosing spline 3 (pink dotted line) implies a considerably larger number of bulk events at low energy than the choice of spline 2 (blue solid line) or spline 1 (red dashed line). As a result, spline 3 leads to a low energy excess of bulk events over the background estimate, shown by the solid black line. This excess is compatible with a DM signal that has a cross-section $\sigma \simeq 9 \times 10^{-42} \ \rm{cm^2}$ (central panel).

For each of these splines, we show the product $\mathcal{P} (d_E | m,\sigma,\mathcal{R}(E))  \mathcal{P}(d_{\mathcal{R}} | \mathcal{R}(E))$ in the right panel. Spline 3 results in a large excess above the background estimate at low-energy (central panel) so its likelihood peaks around $\sigma \simeq 9 \times 10^{-42} \ \rm{cm^2}$ (right panel) indicating a preference for a DM signal with this cross-section.
However, the significance of this result is down-weighted by the prior from the spline fit, since spline 3 gives a poor fit to the rise-time data (left panel). In contrast, spline 2 gives no low-energy excess above background (central panel) and the likelihood, which is largest for vanishingly small cross-sections, shows no preference for a DM signal. Spline 2 provides a good fit to the rise-time data in the left panel resulting in the large posterior probability. Spline 1 results in a small low-energy excess (central panel) but the shallow peak in the right panel indicates that the preference for a DM recoil signal with $\sigma \simeq 3 \times 10^{-42}~\text{cm}^2$ is weak. The posterior distribution is smaller in comparison to spline 2 because the fit to the rise-time data is poorer (left panel).

Marginalising over all splines (including those not shown in the left panel) results in the marginalised posterior (black curve) in the right panel. The absence of any peak (for non-zero values of $\sigma$) shows that there is no preferred value for the DM cross-section after marginalisation. This demonstrates that although a specific choice of $\mathcal{R}(E)$ may give a preference for a DM recoil signal, robustly accounting for all reasonable choices via marginalisation may washout the preferred cross-section picked out by a specific choice.

The results above are for $m = 10$~GeV. We can easily extend our analysis to cover all values of mass $m$. After doing so and extending our marginalisation to over~$40,000$ splines, our marginalised posterior gives a Bayes factor of $\ln\mathcal{B} = -0.69$. This implies a weak preference for the background-only scenario (the interpretation of $\ln\mathcal{B}$ is discussed further in appendix~\ref{sec:numbers}). In addition, we have performed an equivalent frequentist analysis, replacing the marginalisation by a profile likelihood procedure. This results in a p-value of $p = 0.36$, implying that the data prefers the DM signal over the background only hypothesis at less than $1\sigma$.

\subsection{Analysis with the Pareto distribution}
\label{sec:pareto}

As discussed in section~\ref{sec:risetimes}, the bulk population of events may be better described by a Pareto distribution as it gives a better fit to the pulser data at large rise-times (cf.~figure~\ref{fig:lowE_risetimes}).  We now perform the same analysis described in sections~\ref{sec:rt_func} to~\ref{sec:mar} except we model the bulk population with a Pareto distribution rather than a log-normal distribution. This modifies the values of $\mathcal{R}_j$ for each energy bin $E_j$ and so affects the energy-dependence of $\mathcal{R}(E)$.

\begin{figure}[t]
\centering
\includegraphics[width=0.65\textwidth]{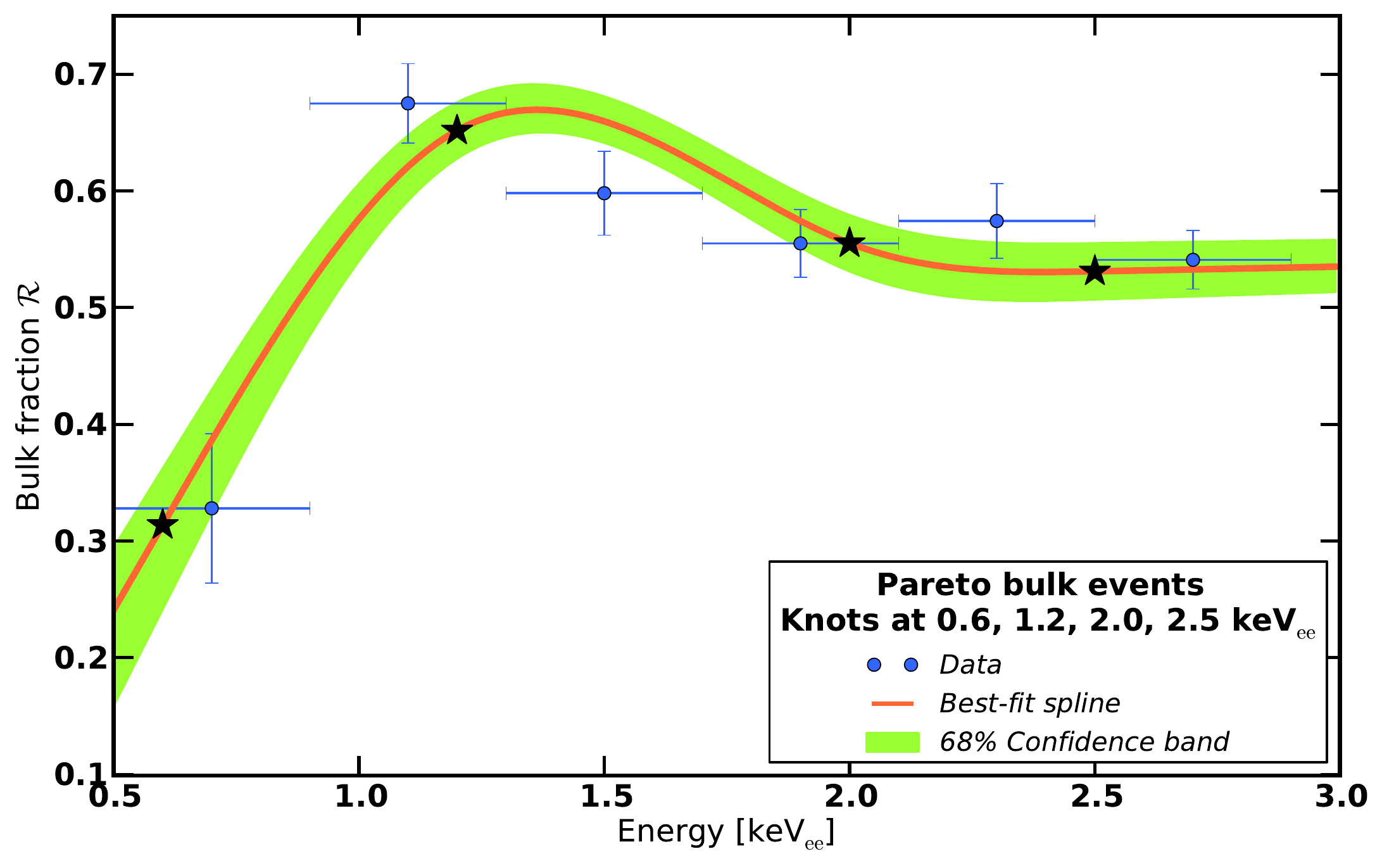}
\caption{The fraction of bulk events $\mathcal{R}$ when fitting the bulk event population with a Pareto distribution. Shown also is the best-fit cubic spline fit (for the knot positions indicated by the black stars) and the $1\sigma$ uncertainty.}
\label{fig:spline_pareto}
\end{figure}

\begin{figure}[t]
\centering
\includegraphics[width=0.99\textwidth]{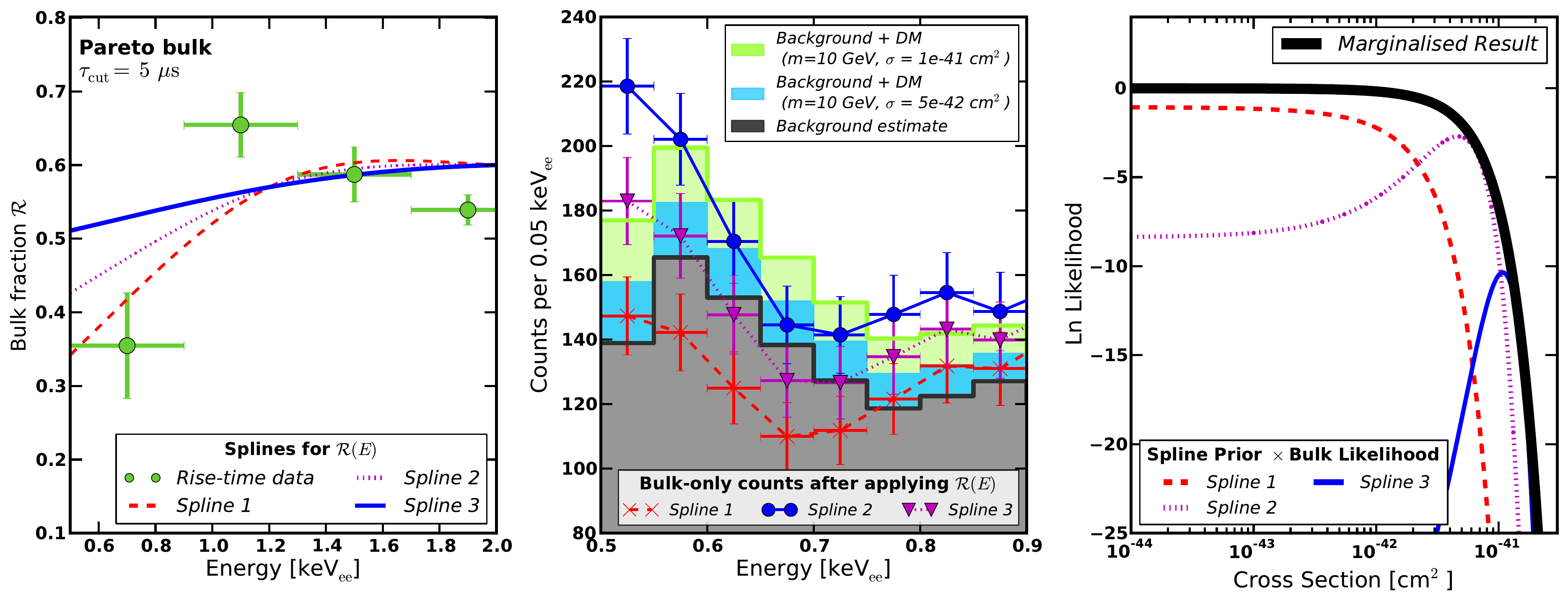}
\caption{Analysis procedure for a 10 GeV DM particle using a Pareto distribution to model the rise-time of the bulk population. The left panel shows three spline choices for the bulk fraction $\mathcal{R}(E)$. The central panel shows the corresponding bulk spectrum for each spline, together with CoGeNT's background estimate (black) and the number of events expected from signal+background (green and blue). Spline 2 and spline 3 result in a low energy excess in the central panel so a non-zero value for~$\sigma$ is preferred, indicated by the peak in the likelihood in the right panel. In contrast, spline 1 results in no low energy excess and no peak in the likelihood. The marginalised posterior (black line, right panel) is flat indicating a vanishing preference for a DM recoil signal over the background estimate.}
\label{fig:likes_pareto}
\end{figure}

The values of $\mathcal{R}$ along with the results of a cubic spline fit are shown in figure~\ref{fig:spline_pareto} (analogous to figure~\ref{fig:bulk_frac} for the log-normal distribution). The slower fall-off of the Pareto distribution means that there are a larger number of bulk events at large rise-times compared to the log-normal distribution. This means that the values of $\mathcal{R}$ are larger than their log-normal counterparts.  However, the overall shape of $\mathcal{R}(E)$, including the drop at low-energy, is similar to that found in figure~\ref{fig:bulk_frac}. 

The result of our analysis when a Pareto distribution is used is shown in figure~\ref{fig:likes_pareto}. Again, although a specific form of $\mathcal{R}(E)$ (such as spline 3) may result in a low-energy spectrum that favours a DM recoil signal, others (such as spline 1) do not. The marginalised posterior function (black solid line in right panel) gives the result including the sum over all splines. As before, the marginalised posterior is flat indicating vanishing evidence for a DM signal. Scanning over all masses results in a Bayes factor of $\ln \, \mathcal{B} = -0.94$ and a p-value of $p = 0.33$. The Bayes factor implies a weak preference for the background only model while the p-value implies that the data prefers the DM signal over the background only hypothesis at less than~$1\sigma$.

\subsection{A smaller rise-time cut}
\label{sec:cuts}

In the previous analysis, we assumed a conservative rise-time cut $\tau_{\rm{cut}}=5~\mu \mathrm{s}$ that meant that the majority of all events were included in our analysis (cf.~figure~\ref{fig:lowE_risetimes}). We now investigate the effect of choosing a smaller value for $\tau_{\rm{cut}}$ in order to directly remove some of the surface events that have a rise-time above this value. This is the approach taken by the CDEX collaboration in their analysis; they choose $\tau_{\mathrm{cut}} = 0.8~\mu \mathrm{s}$~\cite{Yue:2014qdu}.

In the analysis presented in this section we choose $\tau_{\mathrm{cut}} = 0.4~\mu \mathrm{s}$ as this cuts away around~$99.5\%$ of the surface events while leaving around~$40\%$ of the bulk events, assuming the bulk is modelled by a log-normal distribution. The choice for $\tau_{\rm{cut}}$ is somewhat arbitrary but we have tested our analysis with other cuts and reach similar conclusions. 

\begin{figure}[t]
\centering
\includegraphics[width=0.65\textwidth]{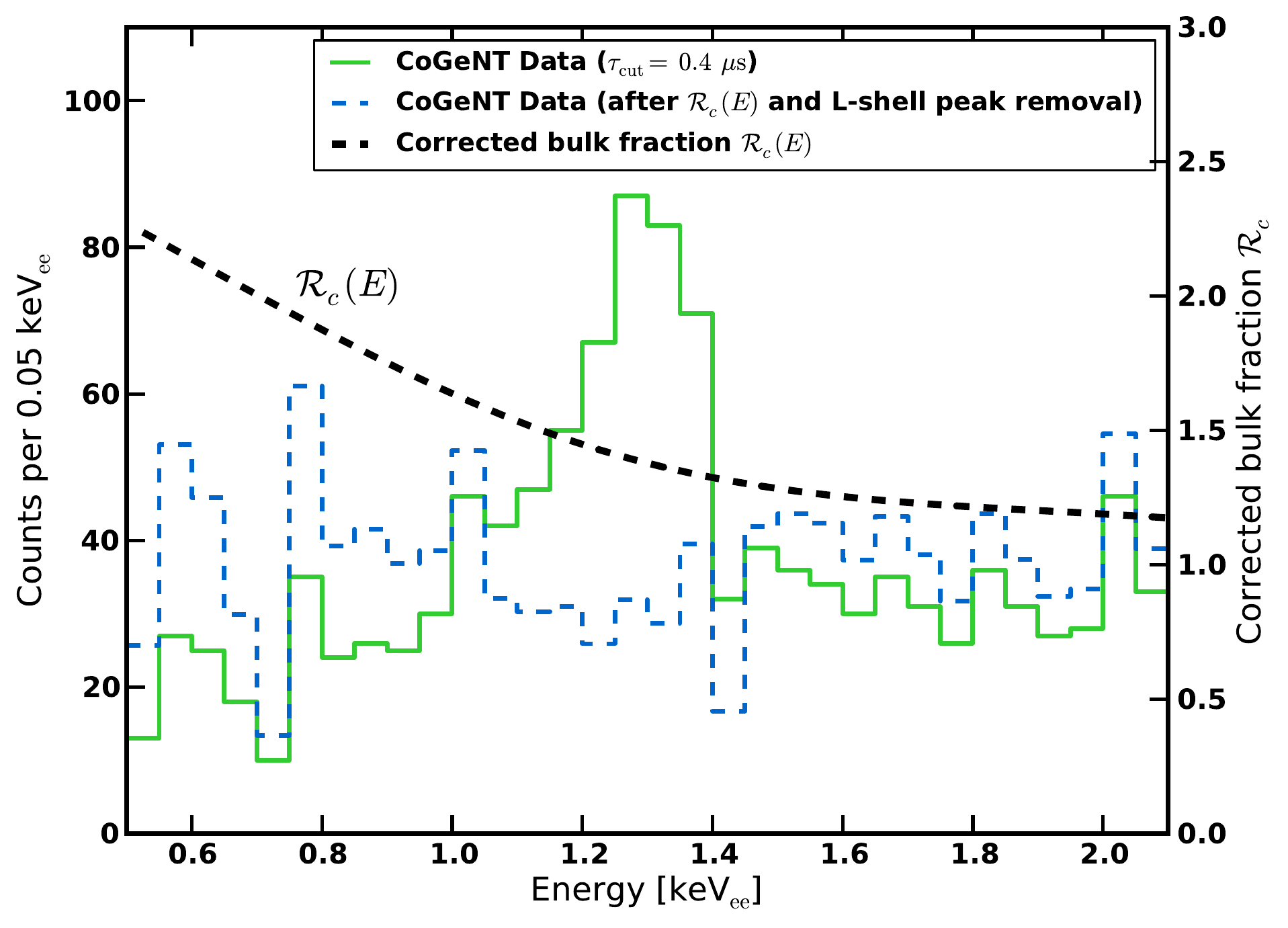}
\caption{Spectrum of CoGeNT events with rise-times below 0.4~$\mu$s. We also show the corrected bulk fraction $\mathcal{R}_{\mathrm{c}}(E)$, which, when applied to the data accounts for surface events as before, and also for bulk events which have been removed by the rise-time cut. }
\label{fig:cut_data}
\end{figure}

\begin{figure}[t]
\centering
\includegraphics[width=0.99\textwidth]{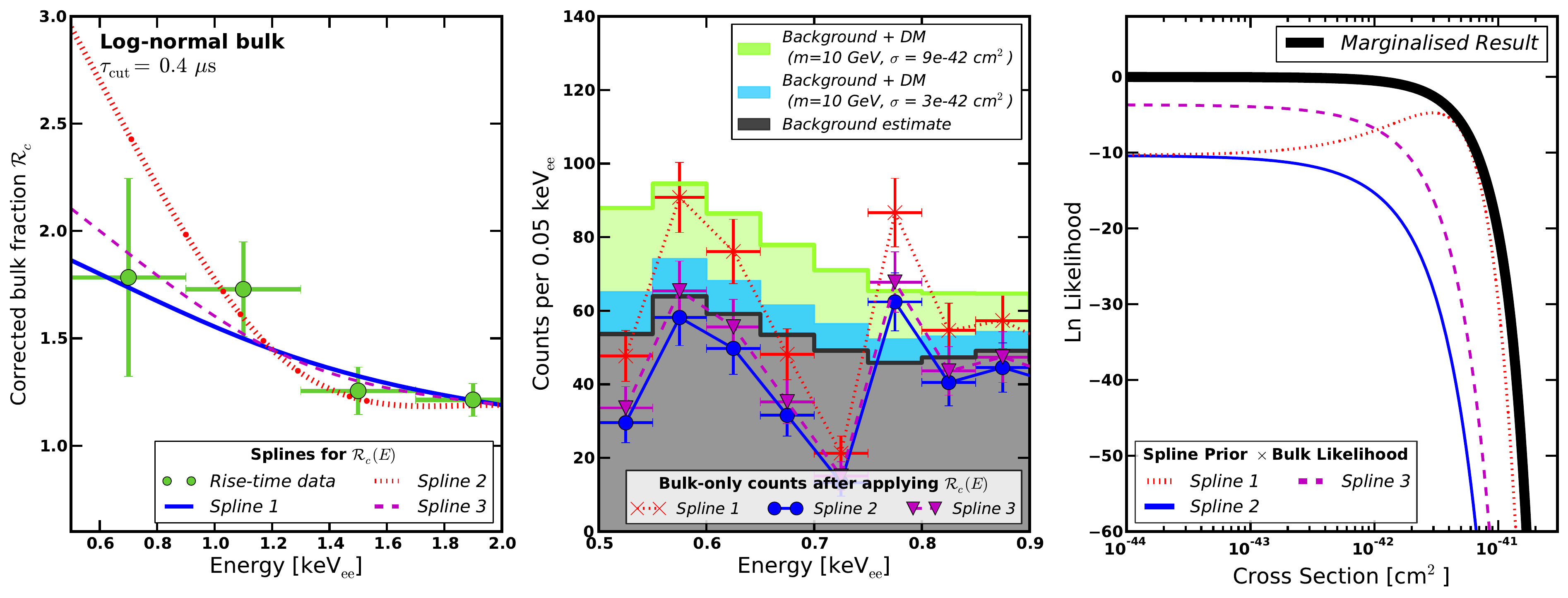}
\caption{Analysis procedure for a 10 GeV DM particle after removing all events with rise-time above $0.4~\mu$s. This cut removes over $99\%$ of the surface events. The left panel shows the corrected bulk fraction data (in green) together with three example spline fits. As can be seen in the central panel, there are no plausible forms of the corrected bulk fraction which fully restore the low-energy excess, giving a featureless marginalised posterior in the right-panel.}
\label{fig:likes_cut04}
\end{figure}

The resulting spectrum after applying the cut \mbox{$\tau_{\mathrm{cut}} = 0.4~\mu \mathrm{s}$} is shown by the solid green curve in figure~\ref{fig:cut_data}. As before, we must correct this spectrum to account for surface events that fall below the timing cut. Additionally, we must also include a cut-efficiency term to correct for the fact that we have removed bulk events. Hence, we define the corrected bulk fraction
\begin{align}
\mathcal{R}_{\mathrm{c}} &= \frac{\mathrm{Num.} \, \mathrm{bulk} < \tau_{\mathrm{cut}}}{\mathrm{Num.} \, \mathrm{total} < \tau_{\mathrm{cut}}} \cdot \frac{\mathrm{Num.} \, \mathrm{bulk} }{\mathrm{Num.} \, \mathrm{bulk} < \tau_{\mathrm{cut}}}\label{eqn:R_corr_1}  \\
&= \frac{\text{Number of bulk events}}{\text{Total number of events}< \tau_{\mathrm{cut}}},
\label{eqn:R_corr}
\end{align}
where the first term in equation~\eqref{eqn:R_corr_1} is the same as equation~\eqref{eqn:Rbin}, while the second term is the cut-efficiency term. We obtain a functional form for $\mathcal{R}_{\mathrm{c}}(E)$ with a cubic spline fit, analogous to the method used to determine $\mathcal{R}(E)$. An example of applying $\mathcal{R}_{\mathrm{c}}(E)$ to the data is shown in figure \ref{fig:cut_data}. In this case, $\mathcal{R}_{\mathrm{c}}(E)$ (black dashed line) is larger than unity because of the cut-efficiency term. The blue dashed line in figure~\ref{fig:cut_data} shows the corrected spectrum (after also subtracting the L-shell peak), which is comparable to the corrected spectrum shown in figure~\ref{fig:cogent_surf_removal} for $\tau_{\rm{cut}}=5~\mu\mathrm{s}$.

Figure~\ref{fig:likes_cut04} shows the results of our analysis when $\tau_{\rm{cut}}=0.4~\mu\mathrm{s}$, for $m=10$~GeV. We assume that the bulk spectrum is modelled by a log-normal distribution. The green data points in the left panel are our results for $\mathcal{R}_{\mathrm{c}}$ and the lines show three example spline fits. The lines in the central panel show the corresponding number of bulk events for each spline choice. The dips at $\sim0.5~\keVee$ and $\sim0.7~\keVee$ are simply a reflection of the reduced number of events measured by CoGeNT at these energies (they are clearly seen in the total spectrum (green line) in figure~\ref{fig:cut_data}). This rise-time cut has effectively removed the vast majority of the surface events and the resulting corrected spectrum shows no clear low energy excess. This is suggestive that the low energy excess arises from surface events. In this case, only the extreme splines (e.g.~spline 2) results in a preference for a DM recoil signal while the marginalised posterior (black line in right panel) is flat, indicating no preference for non-zero~$\sigma$.

After scanning over all DM masses, we obtain a Bayes factor $\mathrm{ln}\, \mathcal{B} = -1.02$ and a p-value of $p = 0.63$. As before, the Bayes factor implies a weak preference for the background only model while the p-factor implies that the data prefers the DM signal over the background at less than $1\sigma$.

\subsection{An energy-dependent rise-time cut}
\label{sec:cuts_Edep}

Instead of the constant rise-time cut used in the previous section, we now use a rise-time cut $\tau_{\mathrm{cut}}$ which varies with energy. This is closer to the procedure followed by the CoGeNT collaboration in~\cite{Aalseth:2010vx,Aalseth:2011wp,Aalseth:2012if,Aalseth:2014eft}. We define our rise-time cut by the time at which the best-fit bulk and surface populations cross. Since the surface events are present predominantly above this cut-off, we will be left with mostly bulk events. For example, for the energy-bin between $0.5$~keV$_{\mathrm{ee}}$ and $0.9$~keV$_{\mathrm{ee}}$, we see from figure \ref{fig:lowE_risetimes} that the distributions cross at $0.68 \, \mu$s (assuming a log-normal for the bulk distribution). This cut removes around $96\%$ of the surface events while maintaining around $80\%$ of the bulk events. Table \ref{tab:cuts} lists all of the rise-time cuts for each energy bin used in our analysis.

\begin{table}[t]\normalsize
\begin{center}
\begin{tabular}{  c  c  }
\toprule                     
  \;Energy [keV$_{\mathrm{ee}}$] &\, Rise-time cut $\tau_{\mathrm{cut}}$ [$\mu$s]  \\   \midrule
    $0.5 \mbox{ - } 0.9$ & 0.68 \\
  $0.9 \mbox{ - } 1.3$ & 0.84 \\
  $1.3 \mbox{ - } 1.7$ & 0.62 \\
  $1.7 \mbox{ - } 2.1$ & 0.64 \\
  $2.1 \mbox{ - } 2.5$ & 0.59 \\
  \bottomrule
\end{tabular}
\end{center}
\caption{The rise-time cut $\tau_{\mathrm{cut}}$ for each energy bin. The cut is defined as the time where the best-fit surface and bulk distributions are equal (assuming a log-normal fit to the bulk events).  All events with rise-times above this value are removed, leaving mostly bulk events only.}
\label{tab:cuts}
\end{table}

\begin{figure}[t]
\centering
\includegraphics[width=0.99\textwidth]{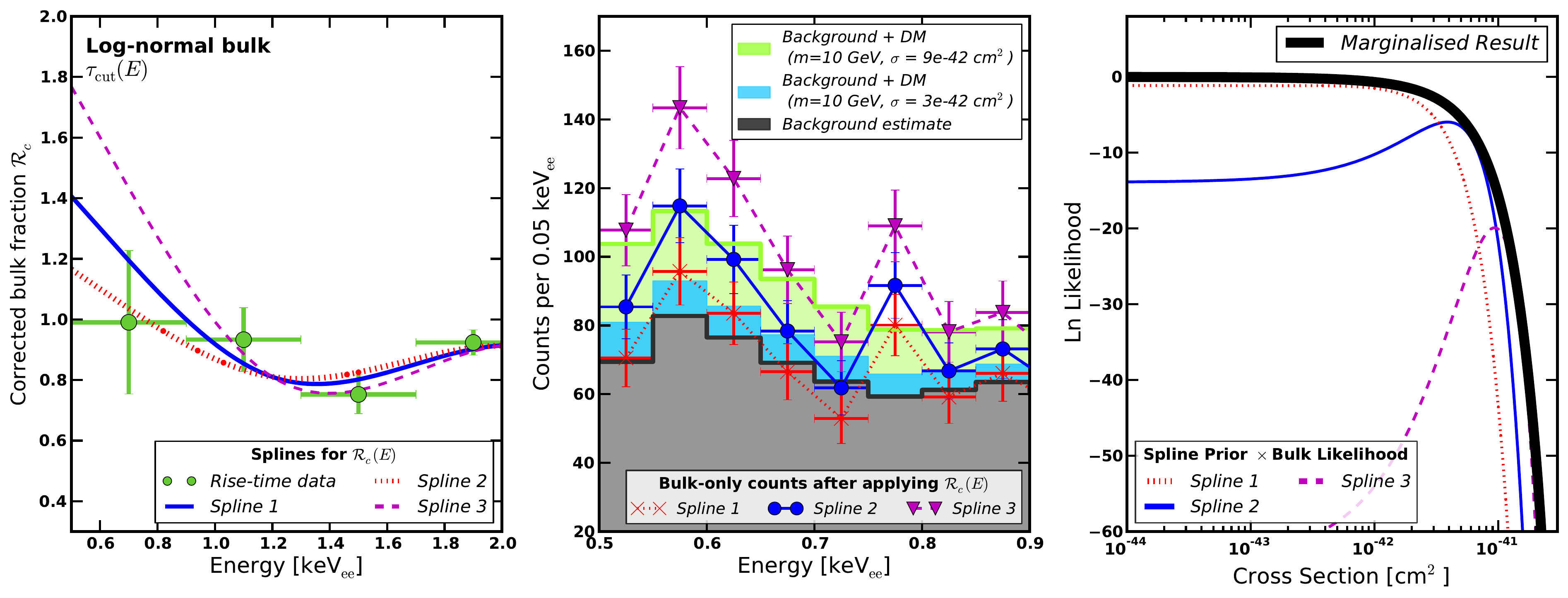}
\caption{Analysis procedure for a 10 GeV DM particles after imposing an energy dependent rise-time cut $\tau_{\mathrm{cut}}(E)$, whose values are listed in table~\ref{tab:cuts}. The left-panel shows the corrected bulk fraction~$\mathcal{R}_{\mathrm{c}}$ together with three different cubic spline fits. These spline fits lead to three different forms for the bulk-only spectrum in the central panel. The presence of an excess of events above background at low energy depends on the choice of spline. After marginalising over all splines, the marginalised posterior (black solid line in right panel) is flat, indicating that the data is fully consistent with known backgrounds. }
\label{fig:likes_cut_Edep}
\end{figure}

The result of our analysis using the energy-dependent rise-time cuts in table~\ref{tab:cuts} are shown in figure~\ref{fig:likes_cut_Edep}. The left panel shows the values of $\mathcal{R}_c$ along with three example spline-fits. These results are in full agreement with our previous findings: splines which give a poor fit to the corrected bulk fraction result in some preference for a DM signal (spline 1 and spline 3) while the better fitting spline results in no preference (spline 2). The absence of any peak in the marginalised posterior indicates that there is no preference for a DM signal when all splines are marginalised over. After including all values of $m$ in our analysis, we obtain a Bayes factor of $\mathrm{ln}\, \mathcal{B} = -0.83$ and a p-value of $p = 0.52$, implying that the data has a weak preference for the background only hypothesis (from the Bayes factor) or that the preference for the DM signal over the background is less than $1\sigma$ (from the p-value).

\section{Analysis of the 807 live-days dataset \label{sec:807}}

All of the analyses in the previous section showed that the preference for a DM signal in the full CoGeNT dataset is non-existent or small; the Bayes factors indicated a slight preference for the background only model while the p-value indicated a preference for DM at less than $1\sigma$ significance. This seems to be at odds with previous results presented by the CoGeNT collaboration, where an irreducible low-energy excess was found even after correcting for the surface event contamination below the rise-time cut. This low-energy excess was interpreted as evidence for DM with mass 7-10~GeV and cross-section $\sim3\times10^{-41}~\text{cm}^2$~\cite{Aalseth:2012if}. In order to show why our results differs from theirs, we also analyse the CoGeNT 807 live-days dataset.

\begin{figure}[t]
\centering
\includegraphics[width=0.65\textwidth]{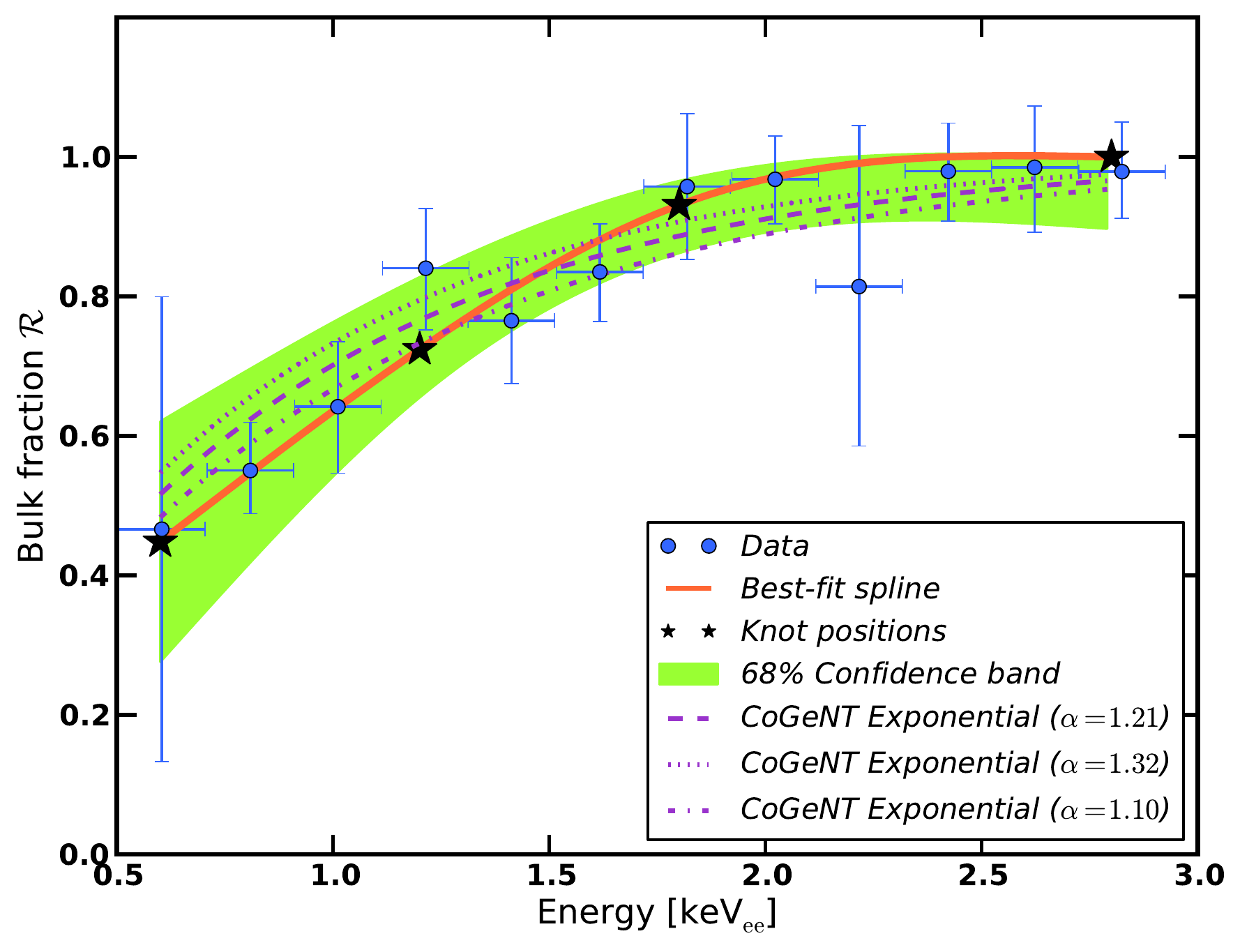}
\caption{Data for the bulk-fraction (from \cite{Aalseth:2012if}). The purple lines show the best-fit exponential used by CoGeNT and its $1\sigma$ error range. Along with this, we show the best-fit cubic spline (red solid) and its associated $68 \%$ uncertainty region (green band).}
\label{fig:spline_807}
\end{figure}

Unlike in previous sections, we now use the bulk fraction $\mathcal{R}$ determined directly by the CoGeNT collaboration in~ref.~\cite{Aalseth:2012if}. We have reproduced these values in figure~\ref{fig:spline_807}. In their analysis, the CoGeNT collaboration employed an energy-dependent rise-time cut that results in a smoother drop-off in the bulk fraction at low energy (cf.~figures~\ref{fig:bulk_frac} and~\ref{fig:spline_807}).

\begin{figure}[t]
\centering
\includegraphics[width=0.99\textwidth]{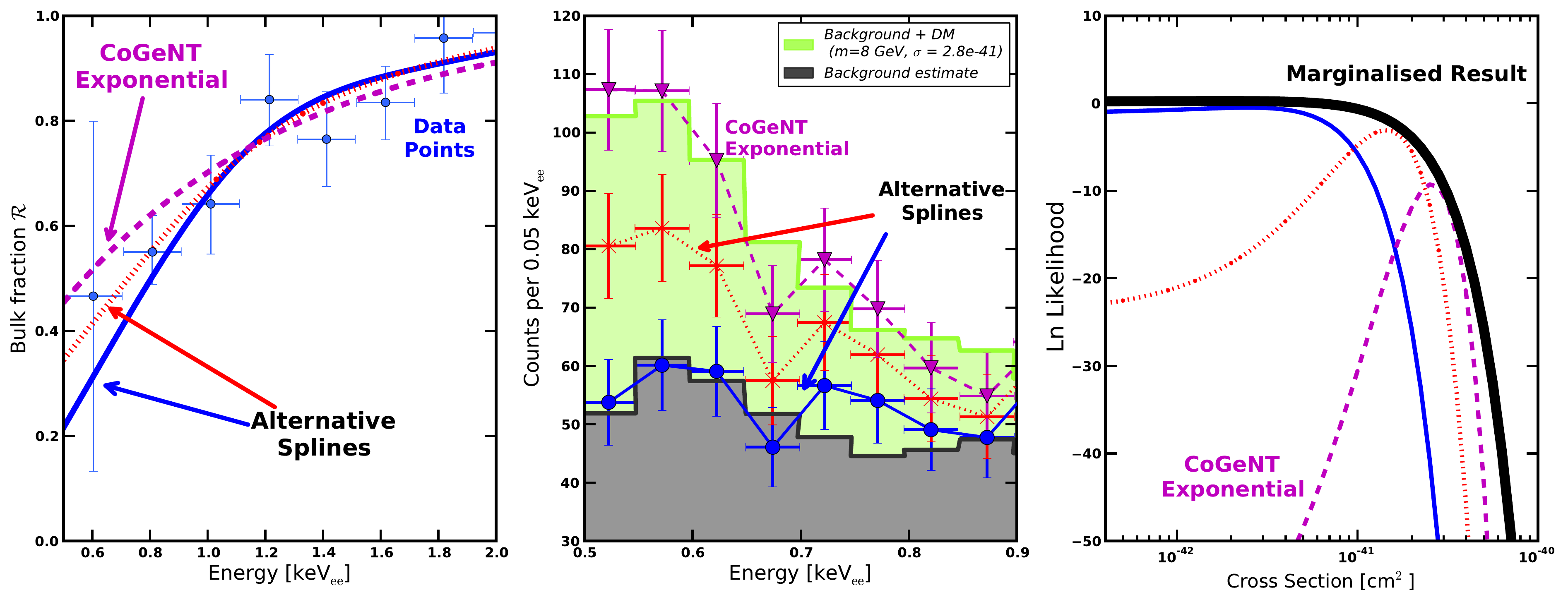}
\caption{Analysis of 807 live-days data for an 8~GeV DM particle. The dashed purple line in the left panel shows the exponential fit used by the CoGeNT collaboration, along with two alternative cubic spline fits to the same data. Both splines fit as well (or better) to this data compared to the exponential. The lines in the central panel show the resulting bulk-only spectrum for each of these functions, together with the background estimate (black) and the number of events expected from a DM signal+background (green). A `low-energy excess' is produced for a limited choice of the fits consistent with $\mathcal{R}$. For the cubic splines, the excess is either reduced or has vanished relative to the CoGeNT exponential (central panel).  The right-hand panel shows that other functional choices give much lower evidence for a DM signal, than the exponential used by CoGeNT. When marginalising over all possible spline choices (black solid line), the signal is washed out leaving vanishing evidence for DM recoils in the CoGeNT 807 live-days data.}
\label{fig:cogent_807}
\end{figure}

To derive their `region of interest', the CoGeNT collaboration fitted a one-parameter exponential of the form $\mathcal{R}_{\mathrm{CoGeNT}}(E) = 1 - \mathrm{exp}[- \alpha \cdot E]$ to the bulk-fraction data $\mathcal{R}$, obtaining $\alpha = 1.21 \pm 0.11$. The purple dashed and dotted lines show the best-fit exponential and its error bars in figure~\ref{fig:spline_807}.  The one-parameter fit is determined by a fit to 12 data points but the fit is dominated by the small error bars at higher-energy. This means that the range of allowed values from the exponential fit is several factors smaller than the errors on the data at the lowest energy, precisely the energy range relevant for an 8~GeV DM particle. As there is no a priori reason to fit the data with an exponential function over another function, we instead use a cubic spline. The result of this fit together with the $1\sigma$ uncertainty is indicated by the green-band in figure~\ref{fig:spline_807}. The cubic spline fit allows the error on $\mathcal{R}(E)$ to grow at lower energies, matching more closely the behaviour of the data points.

To examine the extent to which this choice of exponential function inclines the analysis towards a DM recoil fit, we repeat the analysis of section~\ref{sec:mar} using the 807 live-days data using the bulk-fraction data shown in figure~\ref{fig:spline_807}. We follow the analysis cut scheme described in~\cite{Aalseth:2012if}. The results of the analysis for CoGeNT's exponential fit and two alternative cubic spline fits are shown in figure \ref{fig:cogent_807}.

The central panel shows that the CoGeNT exponential fit results in a low energy excess above the background estimate. This excess is consistent with a DM particle with a mass~8~GeV and cross-section~$\sigma = 2.8 \times 10^{-41}$~cm$^2$. The strongly peaked likelihood in the right panel indicates that there is a strong preference for a DM signal and we obtain the same `region of interest' as the CoGeNT collaboration found in~\cite{Aalseth:2012if} (red region in figure~\ref{fig:cogent_limit}).

Now, the blue solid spline shown in the left panel of figure~\ref{fig:cogent_807} is just as plausible a function for $\mathcal{R}(E)$ (it gives a slightly better fit compared to the CoGeNT exponential). In contrast, this results in no low-energy excess in the central panel and therefore no preference for a DM recoil signal (right panel). As with the 1129 live-days data, we should marginalise over the form of~$\mathcal{R}(E)$ so as not to bias our result by a particular functional form. This result is shown by the solid black line in the right panel of figure~\ref{fig:cogent_807}. It shows that there is no strong preference for a DM recoil over the bulk background, leading to a Bayes factor of $\mathrm{ln} \, \mathcal{B} = -0.76$, consistent with the values we found in section~\ref{sec:1129}.

Hence, we conclude that the CoGeNT `region of interest' is the result of a bias from the choice of the exponential function for $\mathcal{R}(E)$; equally valid functional forms of $\mathcal{R}(E)$ (since there is no theoretically motivated function) result in no excess. The uncertainties from the surface event contamination were not fully accounted for at low-energy when deriving the `region of interest' in~\cite{Aalseth:2012if}.

\section{Summary of results}
\label{sec:sum}

Table~\ref{tab:pvals} summarises the results of our Bayesian and frequentist analyses of the 1129 live-days dataset performed in section~\ref{sec:1129}. In these analyses, we modelled the bulk population by a log-normal and Pareto distribution and considered three variations in the rise-time cut $\tau_{\rm{cut}}$. The results in the `Log-normal' column correspond to the analyses in sections~\ref{sec:mar},~\ref{sec:cuts} and~\ref{sec:cuts_Edep}, where $\tau_{\rm{cut}}=5~\mu$s, $\tau_{\rm{cut}}=0.4~\mu$s and was energy-dependent respectively. The first result in the `Pareto' column corresponds to the analysis in section~\ref{sec:pareto}, where $\tau_{\rm{cut}}=5~\mu$s. For completeness, we also list the results for the Pareto distribution when $\tau_{\rm{cut}}=0.4~\mu$s and when it is energy-dependent. These analyses were not shown in section~\ref{sec:1129} but the methodology is the same as the other analyses.

All of these analyses give consistent results for the p-values and Bayes factor $\mathcal{B}$ from the CoGeNT 1129 live-days dataset. The Bayes factor indicates a slight preference for the background-only model while the p-value indicates a preference for DM at less than $1\sigma$. These results are robust against changes in the bulk background estimate and the bin-size used to calculate $\mathcal{R}(E)$, as we demonstrate in appendices~\ref{sec:bg} and~\ref{app:bin_size} respectively.

\begin{table}[t]\normalsize
\begin{center}
\begin{tabular}{  c  c  c  }
\toprule
& Log-normal & Pareto\\
\midrule
\raisebox{-1.5ex}{$\tau_{\rm{cut}}=5~\mu$s} &  
{$\begin{aligned}[t]
 \ln \mathcal{B}&=-0.69\\
p&=0.36
\end{aligned} $} &
{$\begin{aligned}[t]
 \ln \mathcal{B}&=-0.94\\
p&=0.33
\end{aligned} $}
 \\
  \midrule
  \raisebox{-1.5ex}{$\tau_{\rm{cut}}=0.4~\mu$s} &  
{$\begin{aligned}[t]
 \ln \mathcal{B}&=-1.02\\
p&=0.63
\end{aligned} $} &
{$\begin{aligned}[t]
 \ln \mathcal{B}&=-1.03\\
p&=0.87
\end{aligned} $}
 \\
  \midrule
    \raisebox{-1.5ex}{$\tau_{\rm{cut}}(E)$} &  
{$\begin{aligned}[t]
 \ln \mathcal{B}&=-0.83\\
p&=0.52
\end{aligned} $} &
{$\begin{aligned}[t]
 \ln \mathcal{B}&=-0.68\\
p&=0.45
\end{aligned} $}
 \\
  \bottomrule
  \end{tabular}
\end{center}
\caption{The Bayes factors $\mathcal{B}$ and p-values $p$ from our Bayesian and frequentist profile likelihood marginalisation analyses respectively. These analyses fully incorporate the uncertainties on the bulk fraction $\mathcal{R}(E)$. Lower p-values correspond to a greater preference for a DM signal with $p < 0.32\,(0.05)$ indicating (at least) a $1\sigma\,(2\sigma)$ significance. Bayes factors for which $\ln \mathcal{B} < 0$ imply a preference for the background-only model (i.e.~no DM). Further details are given in appendix~\ref{sec:numbers}.}
\label{tab:pvals}
\end{table}

Given we find no strong preference for a DM signal above background from the CoGeNT data, we proceed to derive a $90\%$ upper limit on the DM-nucleon cross-section. Our limits are defined by integrating under the normalised posterior from $\sigma = 0$ up to the limiting value~$\sigma_{\mathrm{L}}$, such that $90\%$ of the posterior is contained between these two values. 

\begin{figure}[t]
\centering
\includegraphics[width=0.65\textwidth]{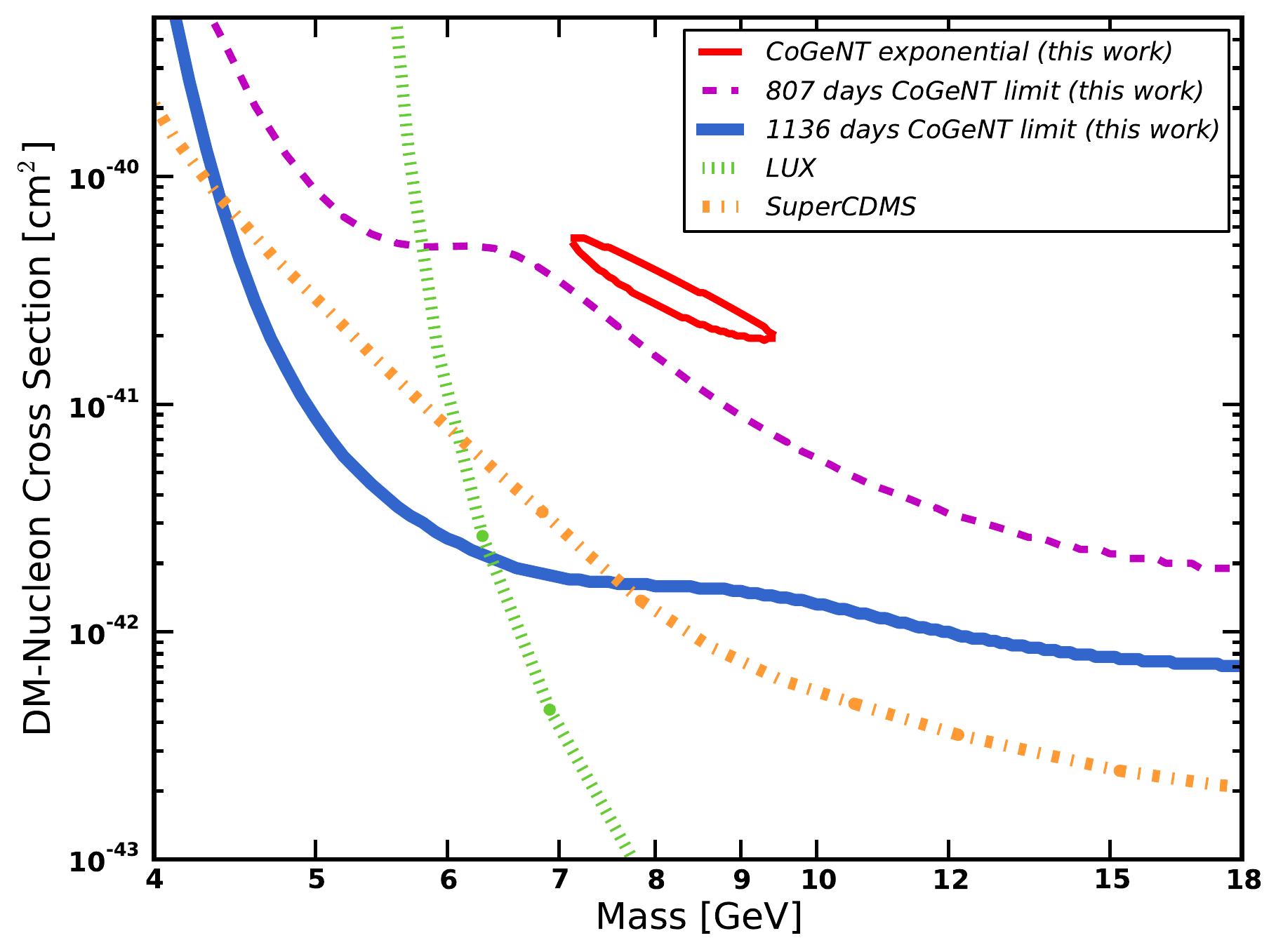}
\caption{The red contour shows the $90\%$ best-fit contour from the 807 live-days data, obtained using CoGeNT's exponential fit to the bulk-fraction, which biases the analysis. The purple line is the $90\%$ limit from the 807 live-days data after marginalising over all cubic spline fits. The blue line is the $90\%$ limit from the full 1129 live-days dataset after marginalising over all cubic splines, and assuming a log-normal distribution for the bulk population and $\tau_{\rm{cut}}=5~\mu$s. Also shown for comparison are the $90\%$ limits from LUX and SuperCDMS. CoGeNT sets the strongest limit in the range 4.5~GeV to 6.5~GeV.}
\label{fig:cogent_limit}
\end{figure}

We first reproduce the earlier CoGeNT `region of interest' from the 807 live-days data, found using their exponential fit for $\mathcal{R}(E)$ (we also marginalise over their quoted uncertainties, which broadens the region slightly). This is shown as the red contour in figure~\ref{fig:cogent_limit}. The purple dashed line shows the $90\%$ upper limit from this data when marginalising over all cubic spline fits to $\mathcal{R}(E)$. 

The solid blue line shows the $90\%$ upper limit from our analysis of the 1129 live-days data after marginalising over all cubic spline fits to $\mathcal{R}(E)$. We model the bulk population with a log-normal distribution and impose $\tau_{\rm{cut}}=5~\mu$s. This limit is competitive with the low-mass limits from the SuperCDMS and LUX experiments (orange and green curves respectively). The 1129 live-days CoGeNT limit is the strongest limit in the narrow window between $4.5$~GeV and $6.5$~GeV.
 
\section{Conclusions}
\label{sec:conc}

We have presented an independent analysis of the most recent 1129 live-days dataset from the CoGeNT experiment. The data contain both bulk and surface events; DM events are dominantly bulk events while the surface events constitute a background which mimics a DM recoil signal at low energy, so should be removed. 

Surface and bulk events can be discriminated by their rise-time (see figure~\ref{fig:lowE_risetimes}). This allows the surface events to be subtracted by using the bulk fraction $\mathcal{R}(E)$ (see figure~\ref{fig:cogent_surf_removal}). Owing to the absence of any theoretically motivated functional form for~$\mathcal{R}(E)$, we parameterised the energy dependence of~$\mathcal{R}(E)$ with cubic splines (see figure~\ref{fig:bulk_frac}). Using these cubic splines, we found that an excess above background arises only for specific choices of the bulk fraction~$\mathcal{R}(E)$, while other equally plausible spline choices give no excess. We demonstrated this repeatedly in section~\ref{sec:1129} (see e.g.~figure~\ref{fig:mar_fig}). A robust (unbiased) analysis should therefore be independent of the choice for $\mathcal{R}(E)$. We accomplished this by marginalising over all splines; we integrated out $\mathcal{R}(E)$ using Bayesian and frequentist methods and found less than~$1\sigma$ evidence for a DM recoil signal in CoGeNT data, above the known backgrounds.

To test the robustness of our results, we demonstrated that we obtain the same conclusions under different assumptions. We first considered an alternative model for the rise-time of the bulk events. The Pareto distribution provides a better fit to pulser data at high rise-times (see figure~\ref{fig:lowE_risetimes}) but the resulting preference for DM from the data is still less than~$1\sigma$. This is shown graphically in figure~\ref{fig:likes_pareto}. Additionally, we considered variations in the rise-time cut: an energy-independent cut at 0.4~$\mu$s that removes the vast majority of surface events and an energy-dependent cut. In both cases, we found that DM is preferred over background at less than~$1\sigma$ significance (see figures~\ref{fig:likes_cut04} and~\ref{fig:likes_cut_Edep}). 

Furthermore, we have analysed the 807 live-days data that was used by the CoGeNT collaboration to define a `region of interest' in the plane of DM mass and cross-section~\cite{Aalseth:2012if}. We showed that this region of interest is the result of a bias coming from the choice of the exponential functional form for~$\mathcal{R}(E)$. We demonstrated in figure~\ref{fig:cogent_807} that other equally plausible functional choices give a reduced (or zero) preference for DM recoils. We found that the region of interest vanishes for the marginalised result, which accounts for all reasonable choices of $\mathcal{R}(E)$.

Thus far, we have not commented on the $2.2\sigma$ significance for an annual modulation also present in this dataset~\cite{Aalseth:2014eft}. We note only that if this annual modulation is attributed to DM recoils, then the apparent lack of consistency with the unmodulated data must be explained (see~\cite{HerreroGarcia:2011aa}).

In conclusion, we find that there is at most a~$1\sigma$ evidence for a DM recoil signal in CoGeNT data. Using the same dataset to derive a limit on the DM-nucleon scattering cross-section, we found that the CoGeNT limit is competitive with those from other experiments (see figure~\ref{fig:cogent_limit}) and sets the strongest limit in the mass range $4.5$~GeV to $6.5$~GeV.

\acknowledgments{
We thank the CoGeNT collaboration for publicly releasing their dataset and Juan Collar for encouraging us to consider distributions other than the log-normal distribution. CM thanks Johnathan McCabe for discussions regarding distribution functions. JHD gratefully acknowledges the STFC for financial support. CB thanks CERN for hospitality while part of this work was carried out.
}

\appendix
\section{Log-normal and Pareto distributions}
\label{app:dists}

Assuming that the bulk and surface event populations follow a log-normal distribution, we fit the following six-parameter function to the rise-time $\tau$ data (e.g.~as shown in figure~\ref{fig:lowE_risetimes})
\begin{align}
f_{\mathrm{total}}(\tau) &= f_{\mathrm{bulk}}(\tau) +  f_{\mathrm{surface}}(\tau) \nonumber \\
 &= \frac{1}{\tau \sqrt{2 \pi}} \left( \frac{A_b}{\sigma_b} \mathrm{exp}\left[ -\frac{(\mathrm{ln} \tau - \mu_b)^2}{2 \sigma_b^2} \right]  + \frac{A_s}{\sigma_s} \mathrm{exp}\left[ -\frac{(\mathrm{ln} \tau - \mu_s)^2}{2 \sigma_s^2} \right] \right), \nonumber
\label{eqn:log_normal}
\end{align}
where $A_b$ and $A_s$ are the amplitudes of the bulk and surface distributions respectively, $\mu_b$ and $\mu_s$ are the location parameters and $\sigma_b$ and $\sigma_s$ are the scale parameters.

Alternatively, the bulk events can be modelled with a Type IV Pareto distribution (with location parameter $\mu = 0$). In that case, the bulk population is modelled with
\begin{equation}
f_{\mathrm{bulk}}(\tau)=\frac{A_b \alpha_b }{\gamma_b \kappa_b^{1/\gamma_b}}\tau^{1/\gamma_b - 1} \left[  \left( \frac{\kappa_b}{\tau} \right)^{-1/\gamma_b} +1 \right]^{-\alpha_b-1} \;.
\end{equation}
In section~\ref{sec:1129} we fixed $\alpha_b = 0.2$ so that the bulk distribution has three free parameters: the amplitude $A_b$ and the variables $\gamma_b$ and $\kappa_b$. The distribution shown in figure~\ref{fig:lowE_risetimes} has $\{A_b,\gamma_b,\kappa_b\}=\{138,0.249,0.300\}$. 

\begin{figure}[t]
\centering
\includegraphics[width=0.55\textwidth]{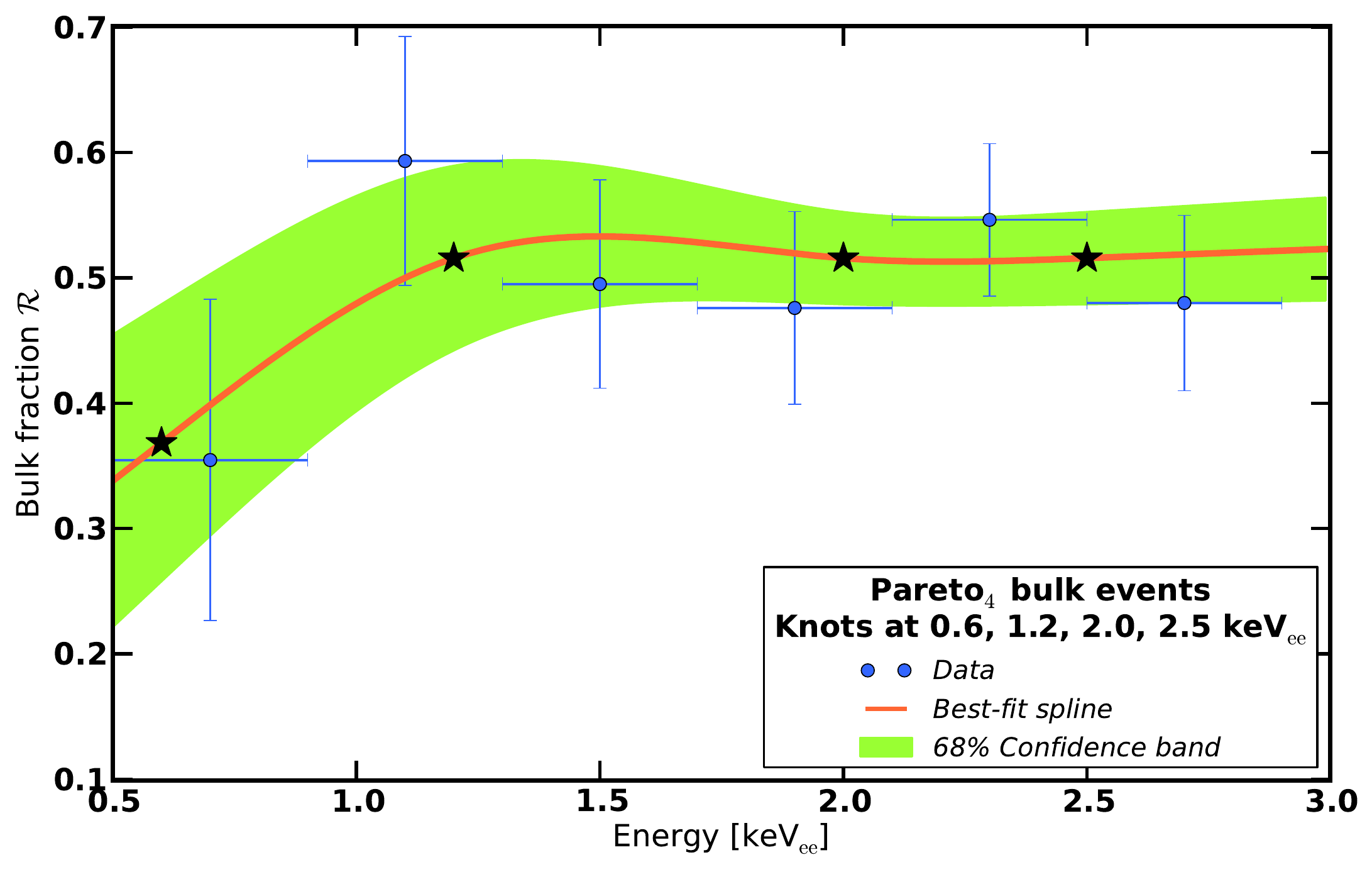}
\caption{Values of the bulk fraction $\mathcal{R}$ obtained by fitting the bulk population to the Pareto$_4$ distribution. This distribution has an additional free parameter compared to the analyses in section~\ref{sec:1129}. The best-fit spline and its $1\sigma$ uncertainties are also shown.}
\label{fig:Pareto4_spline}
\end{figure}

\begin{figure}[t]
\centering
\includegraphics[width=0.99\textwidth]{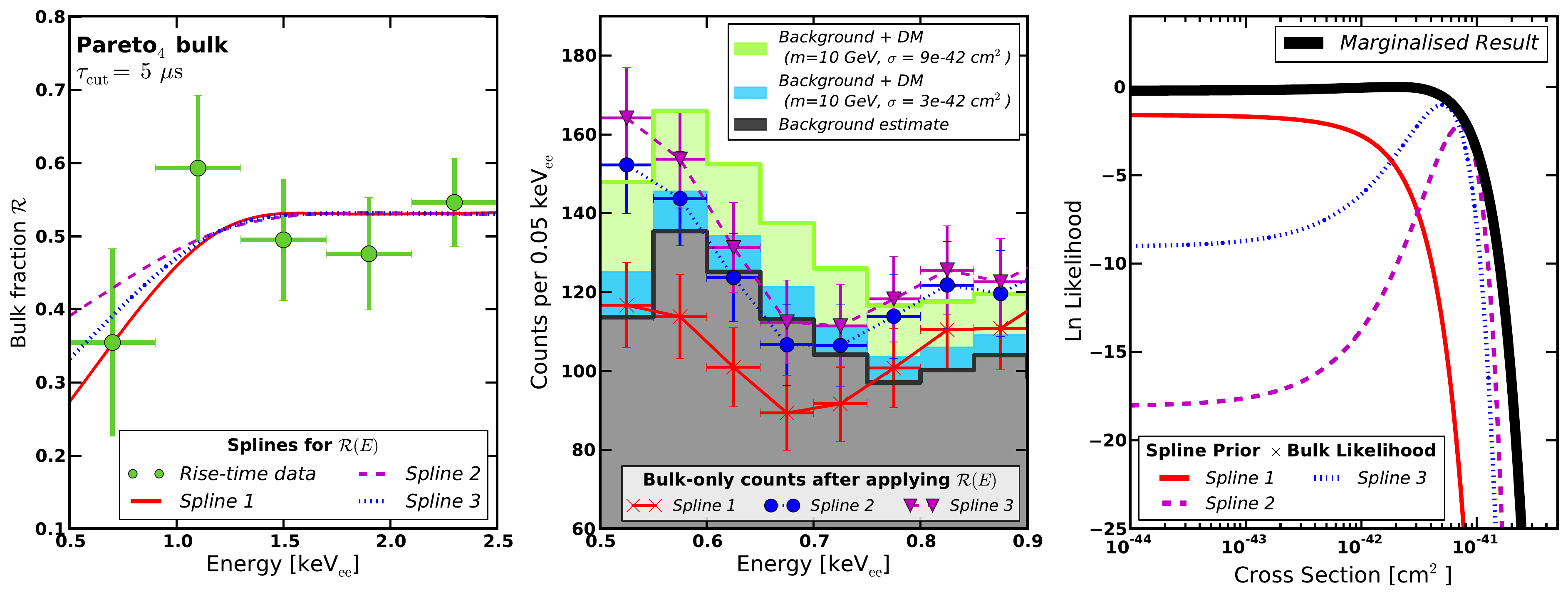}
\caption{Analysis for a $10$~GeV DM particle when the bulk event rise-times are modelled with a Pareto$_4$ distribution. The extra parameter in the Pareto$_4$ distribution results in larger error bars on~$\mathcal{R}$ (left-panel). Owing to this, the priors for each spline shown in the left panel are of similar size. This means that spline 2 and 3 make a greater contribution to the marginalised posterior than in the analyses in section~\ref{sec:1129}, which results in a very small peak in the marginalised posterior. Despite this, the data prefer a DM signal at less than $1.2\sigma$ above the background-only scenario.}
\label{fig:likes_pareto4}
\end{figure}

We constrained the parameter $\alpha_b$ so that the log-normal and Pareto fits were performed with the same number of free parameters. To show that the choice for $\alpha_b$ does not bias our result, we here show the result when $\alpha_b$ is also allowed to vary. We call this distribution Pareto$_4$ as there are now four free-parameters. In this case the fit to the rise-time data has seven free parameters: four from the Pareto$_4$ distribution for the bulk events and three from the log-normal distribution for the surface events. 

Values for the bulk fraction~$\mathcal{R}$ are shown in figure \ref{fig:Pareto4_spline}, along with the best-fit cubic spline and its $1\sigma$ uncertainty band. Owing to the extra parameter, the uncertainties on~$\mathcal{R}$ are larger for each bin. This is reflected in the $1\sigma$ band from our cubic spline fits, which is wider. We show in figure~\ref{fig:likes_pareto4} the analysis with the Pareto$_4$ distribution and $\tau_{\rm{cut}}=5~\mu$s. From this analysis we obtain a p-value of $p = 0.23$ and a Bayes factor of $\mathrm{ln} \, \mathcal{B} = -0.49$, after marginalising over all splines (including allowing the energy values of the first two knots to vary). These results are consistent with those found earlier: DM is preferred above background at less than $1.2\sigma$.

\section{Sources of bulk events}

In this section, we discuss the spectra used in our analysis for the bulk event background and the DM recoil signal.

\subsection{Cosmogenic and radioactive backgrounds}
\label{sec:bg}

In addition to the surface events, there are a few known backgrounds expected in the bulk of the detector. The L and K shell peaks, which appear prominently in the CoGeNT spectrum, arise from activated cosmogenic isotopes which decay to give lines at various (known) energies. By fitting to the K-shell, we can subtract the L-shell events at lower energy \cite{Aalseth:2014jpa,Aalseth:2012if}. 

The K-shell peaks are all present above 4 keV$_{\mathrm{ee}}$. The fit proceeds as follows:
we fit a gaussian to the K-shell peak for each isotope given in~\cite{Aalseth:2011wp,Aalseth:2014eft} with the mean and width fixed and the amplitude allowed to vary freely. There is additionally a roughly constant component in the fit, which is mostly flat but falls gradually towards higher energies \cite{Aalseth:2014jpa}. This gives us the amplitudes of each K-shell peak. We use the ratios of the amplitudes of the L and K shell peaks for each isotope~\cite{Aalseth:2014eft} to build their L-shell peaks, which are all present around 1.3~keV$_{\mathrm{ee}}$. We then subtract these from the spectrum, as shown in figure~\ref{fig:cogent_surf_removal}.\footnote{The CoGeNT collaboration have also noted that the amplitude of the L-shell peak requires a $10\%$ correction factor  \cite{Aalseth:2011wp,Aalseth:2012if,Aalseth:2014eft}. We have ensured that our conclusions are unchanged with and without this factor, and have also performed our analysis while marginalising over the amplitude of the L-shell peak.} 

CoGeNT also discuss other backgrounds. The dominant contributions for bulk events arise from muons in the shielding of the detector, radioactive isotopes in the resistors and $\alpha$ and neutron emission from the cavern walls \cite{Aalseth:2012if}. For these events, we use the bulk background spectrum derived in~\cite{Aalseth:2012if}, corrected for the increased exposure. It is clear that there is a flat plateau in this spectrum, but these events also extend into the signal region, with a small rise at low energy. For our analysis, the amplitude of the bulk background is allowed to float, which in practice means that its amplitude is fixed to match the (flat) bulk spectrum at energies above the L-shell peak.

\begin{figure}[t]
\centering
\includegraphics[width=0.99\textwidth]{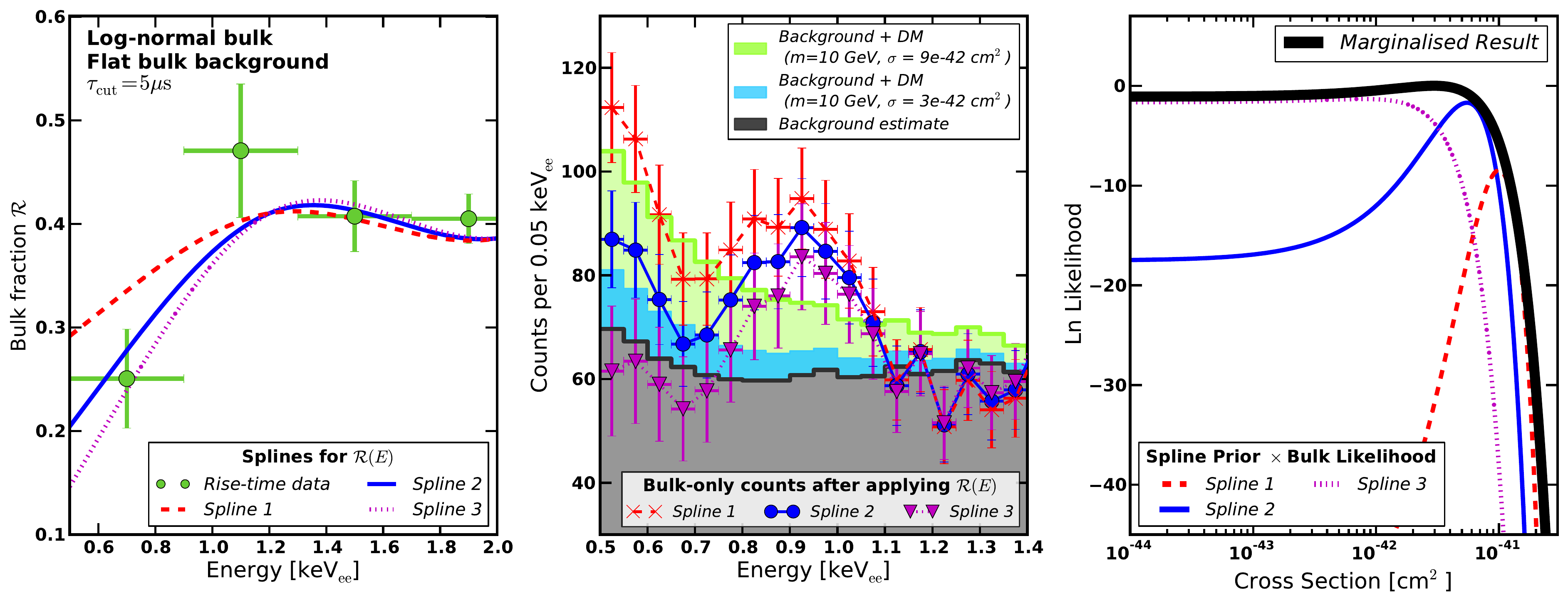}
\caption{Analysis of 1129 live-days of CoGeNT data with the bulk background estimate from~\cite{Aalseth:2014jpa}. In this case, the blue solid spline now gives strong evidence for a light DM recoil signal. However, as can be seen in the right panel, the result after marginalising is similar to that in figure~\ref{fig:mar_fig}, since the purple dashed spline is fully consistent with the bulk background estimate. Hence, our marginalised result is robust to changes in the bulk background.}
\label{fig:likes_flatbg}
\end{figure}

We have also cross-checked our results using the bulk backgrounds discussed in~\cite{Aalseth:2014jpa}, and using a flat background. For example, we show in figure~\ref{fig:likes_flatbg} the result of our analysis performed using only the flat component from~\cite{Aalseth:2014jpa} for the bulk background estimate. Comparing with figure~\ref{fig:mar_fig}, we see that in this case the solid blue spline now gives strong evidence for a low-energy excess, consistent with a light-DM recoil. However, the purple dashed spline, which is a better fit to the data for~$\mathcal{R}$ in the left panel, gives no low-energy excess and has no preference for a DM recoil signal.

Changing the bulk background estimate has altered the fit for each individual spline, but the result is the similar to that found in section~\ref{sec:mar} after marginalising over all possible splines (solid black line in right panel). This demonstrates that our result is robust against reasonable changes in the bulk background spectrum.

\subsection{Dark matter}
\label{sec:recoil}

We generate the DM recoil spectrum $ f(m,\sigma)$ using the following expression,
\begin{equation}
\label{eqn:recoil_rate}
f(m,\sigma) \equiv \frac{\mathrm{d}R}{\mathrm{d}E} = \epsilon(E_{\mathrm{nr}}) \frac{\sigma (E_{\mathrm{nr}})}{2 m \mu^2} \rho \eta (E_{\mathrm{nr}}) ,
\end{equation}
where $\sigma(E_{\mathrm{nr}})$ is the DM-nucleus cross-section as a function of nuclear-recoil energy $E_{\mathrm{nr}}$, $\epsilon$ is an efficiency factor accounting for the CoGeNT cut-acceptance \cite{PhysRevLett.106.131301,Aalseth:2014eft2} (for the 807 live-days dataset we follow the cut scheme in~\cite{Aalseth:2012if}), $\mu$ is the DM-nucleus reduced mass, $\rho = 0.3 \, \mathrm{GeVcm}^{-3}$ is the local dark matter density and $\eta(E_{\mathrm{nr}}) = \int_{v_{\mathrm{min}}}^{\infty} \mathrm{d}^3 v \frac{f(v + u_{\mathrm{e}})}{v} $ is the DM mean velocity. The mean velocity is integrated over the distribution of DM particle velocities in the galaxy $f(v)$ boosted into the reference frame of the Earth by $u_{\mathrm{e}}$~\cite{McCabe:2013kea,Lee:2013xxa}. The lower limit of the integration is $v_{\mathrm{min}} (E_{\mathrm{nr}})$, which is the minimum DM velocity required to induce a recoil of energy $E_{\mathrm{nr}}$. We assume the standard halo model such that $f(v)$ is given by a Maxwell-Boltzmann distribution cut off at an escape velocity of $v_{\mathrm{esc}} = 537 \, \mathrm{kms}^{-1}$ \cite{Piffl:2013mla}; in this case, there is an analytic form for $\eta(E_{\mathrm{nr}})$~\cite{McCabe:2010zh}.

We assume that DM particles interact identically with protons and neutrons so that $\sigma(E_{\mathrm{nr}}) = \sigma \left( \mu/\mu_p \right)^2 A^2 F^2(E_{\mathrm{nr}})$, where $\sigma$ is the zero-momentum DM-nucleon cross-section, $A$ is the atomic mass of germanium, $\mu_p$ is the DM-proton reduced mass and $F(E_{\mathrm{nr}})$ is the Helm nuclear form factor~\cite{Lewin199687}. We convert nuclear-recoil energy $E_{\mathrm{nr}}$ into electron-equivalent energy $E$ using the relation $E= 0.2 E_{\mathrm{nr}}^{1.12}$~\cite{Aalseth:2012if}. 

\section{Variations in the bulk fraction $\mathcal{R}$ bin-size}
\label{app:bin_size}

\begin{figure}[t]
\centering
\includegraphics[width=0.99\textwidth]{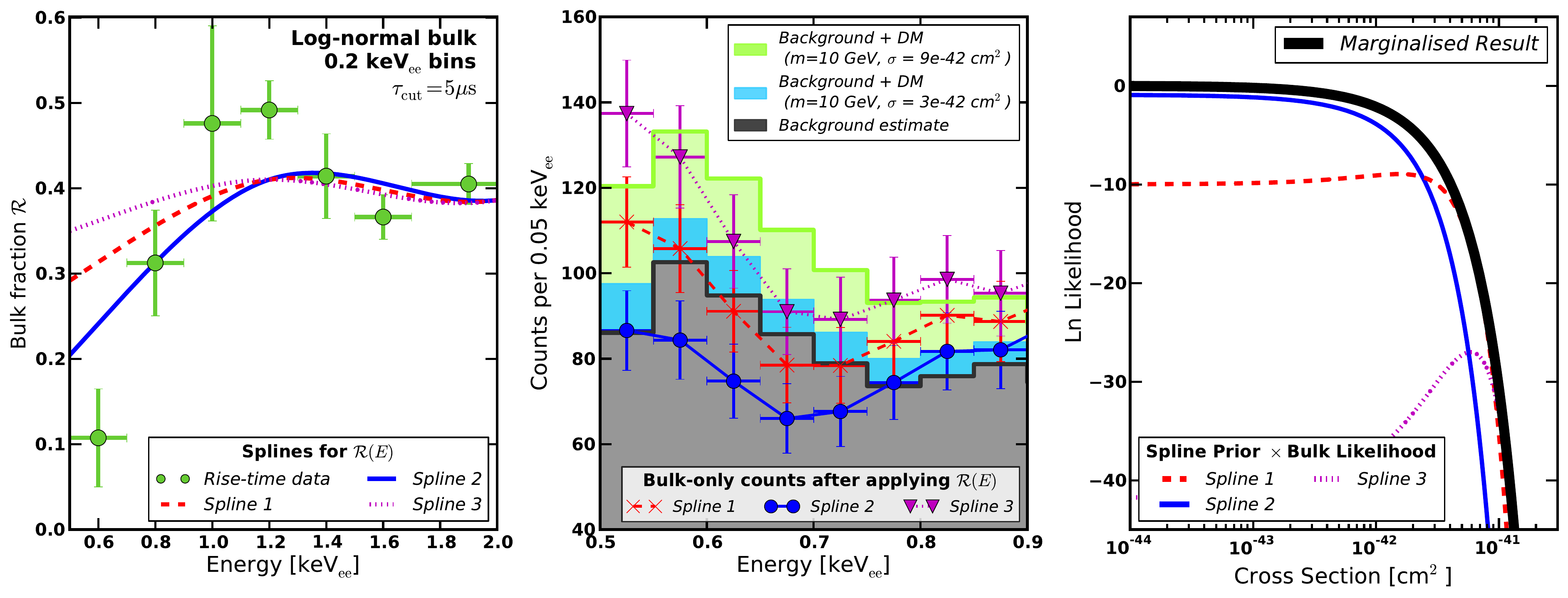}
\caption{Analysis of 1129 live-days of CoGeNT data when the bin size used to calculate $\mathcal{R}$ is~$0.2$ keV$_{\mathrm{ee}}$. In this case, the fall in the bulk fraction at low-energy is even more pronounced (cf.~figure~\ref{fig:mar_fig}). This results in splines 1 and 3 being strongly disfavoured, leading to their priors being further suppressed. Hence, the CoGeNT data is even less compatible with a DM recoil signal since the splines which result in a low-energy excess do not fit well to the bulk fraction data (left panel).}
\label{fig:likes_02}
\end{figure}

To calculate the bulk fraction~$\mathcal{R}$, we separated the data into bins of size $0.4$ keV$_{\mathrm{ee}}$. The choice of bin size is a compromise between resolution in energy and the error in each value of~$\mathcal{R}$ (a larger bin size leads to smaller errors). As this choice is arbitrary, we demonstrate here that our conclusions remain unchanged with a different bin size.

We choose instead to bin the data between $0.5$ keV$_{\mathrm{ee}}$ and $1.7$ keV$_{\mathrm{ee}}$ in bins of size $0.2$ keV$_{\mathrm{ee}}$, while keeping the bin size of the larger bins at higher energy fixed ($\mathcal{R}$ is largely flat at these energies so the bin size is less important.). The results of calculating $\mathcal{R}$ for these smaller bins is shown in the left panel of figure~\ref{fig:likes_02}, along with our cubic spline fits. We see from the figure that with the smaller bins, the drop in bulk fraction at low energies is now even clearer than for the larger bins used in figure \ref{fig:bulk_frac}.  

We show the impact this has on our analysis in figure \ref{fig:likes_02}. Owing to the more rapid drop at low energy, splines 1 and 3 are now further disfavoured, which means their prior values are suppressed to a greater degree than for the case considered in section~\ref{sec:mar}. This can be seen in the right-panel, where the product $\mathcal{P} (d_E | m,\sigma,\mathcal{R})  \mathcal{P}(d_{\mathcal{R}} | \mathcal{R})$ is now even more suppressed for splines 1 and 3 while even spline 2 gives too many events at low-energy. 

Hence, using smaller energy bins to calculate $\mathcal{R}$ further disfavours a DM interpretation of the CoGeNT data. Now the splines that result in a low-energy excess give an even worse fit to the data for the bulk fraction. We therefore find that our conclusions are robust against decreasing the size of the energy-bins used to calculate $\mathcal{R}(E)$.

\section{Variations in the energy range in our likelihood analysis}
\label{sec:energy_range}

\begin{figure}[t]
\centering
\includegraphics[width=0.99\textwidth]{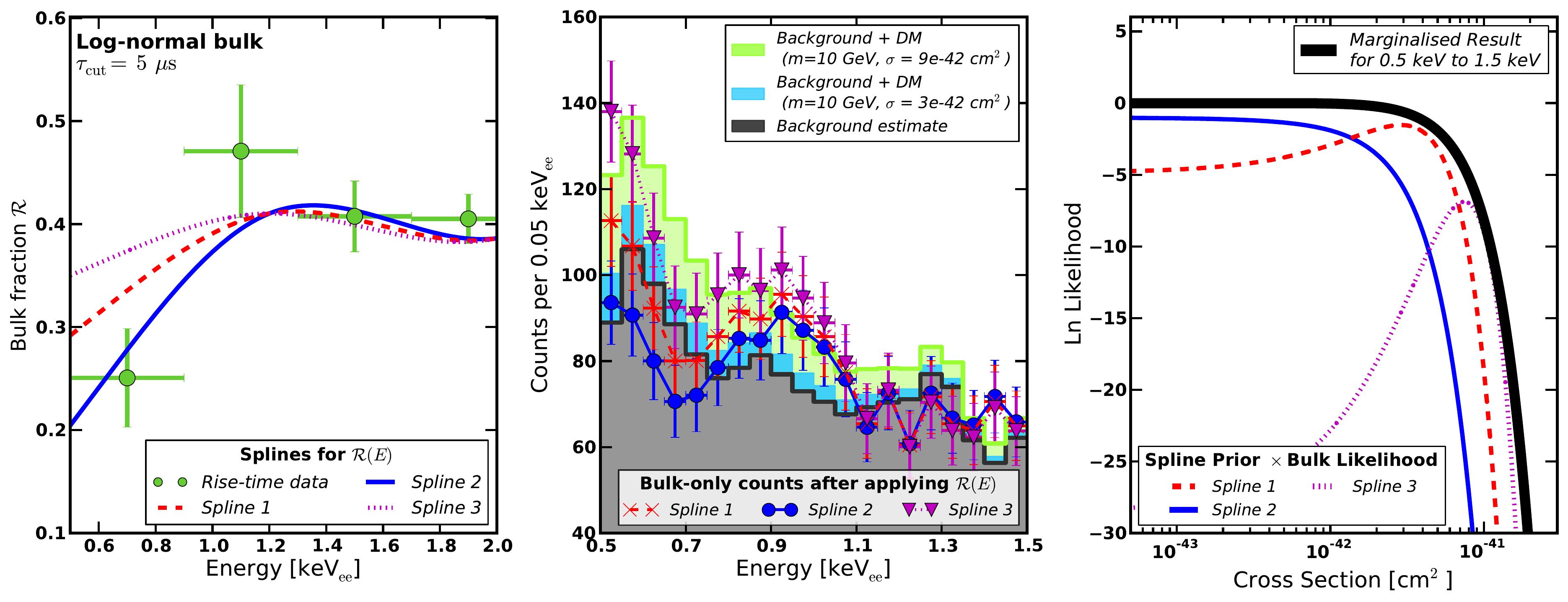}
\includegraphics[width=0.99\textwidth]{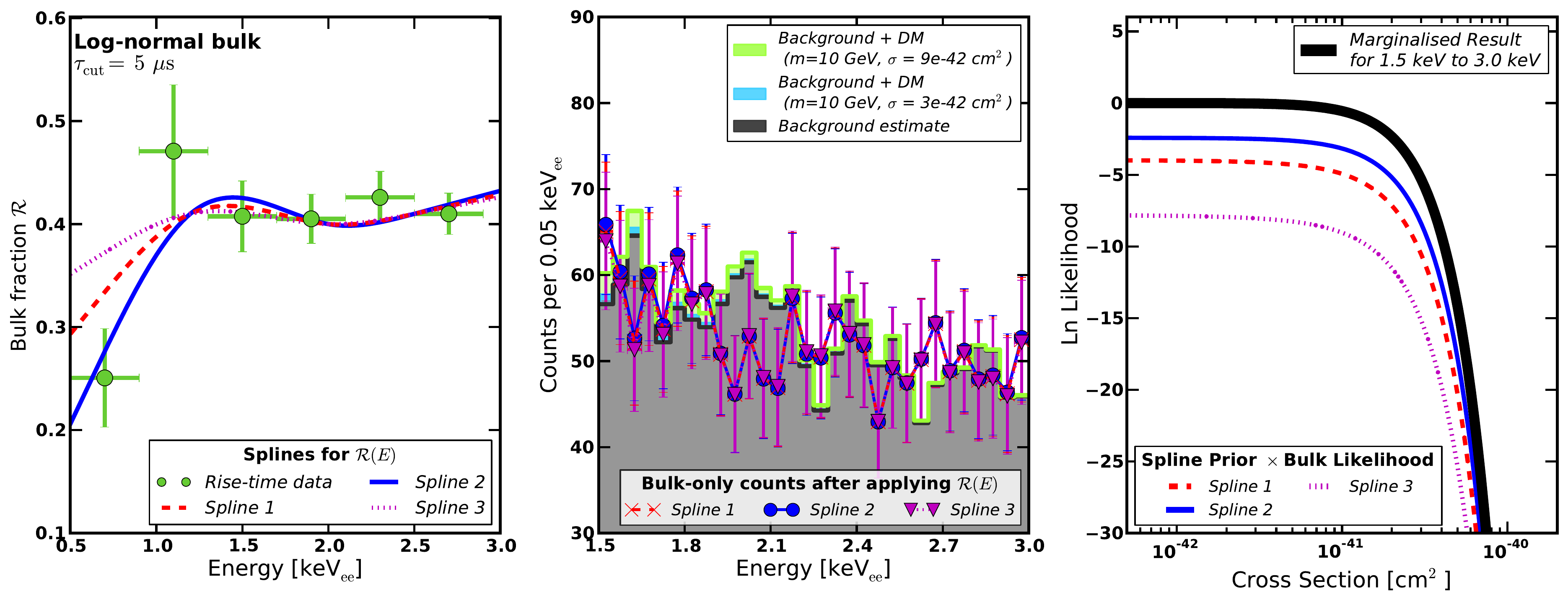}
\caption{Analysis of 1129 live-days of CoGeNT data for a 10 GeV DM particle for different energy ranges in the likelihood analysis. The upper (lower) panels show the results for the range 0.5 to $1.5~\keVee$ (1.5 to $3.0~\keVee$). The upper panels show that the energy range 0.5 to $1.5~\keVee$ gives similar results to the full range 0.5 to $3.0~\keVee$ used in section~\ref{sec:1129} (cf.~figure~\ref{fig:mar_fig}). The lower panels show that there is no preference for DM in the range 1.5 to $3.0~\keVee$ for any choice of spline.}
\label{fig:likes_energy}
\end{figure}

We used the energy range 0.5 to~$3.0~\keVee$ in our likelihood analysis in section~\ref{sec:1129}. In this appendix we investigate the effect of two variations in this range on our likelihood analysis.

Firstly, we consider the restricted range 0.5 to~$1.5~\keVee$. We would expect all of the DM signal to be in this range if the DM scatters elastically with the usual spin-independent interaction and has a mass around 10~GeV. The results of this analysis are shown in the upper panels in figure~\ref{fig:likes_energy}. In this case, we still use the range 0.5 to $3.0~\keVee$ to determine the bulk fraction $\mathcal{R}(E)$ but only consider the range 0.5 to $1.5~\keVee$ when determining the bulk likelihood. In this instance, splines 1 and 3 result in slightly higher likelihoods (cf.~figure~\ref{fig:mar_fig}) but the marginalised result is similar to previous results; after marginalising over all spines we obtain a p-value of $p = 0.43$ and a Bayes factor of $\mathrm{ln} \, \mathcal{B} = -0.99$, consistent with the values in table~\ref{tab:pvals}.

Secondly, we consider the high energy range 1.5 to~$3.0~\keVee$. While a DM signal would not be present in this range if it scatters elastically with the usual spin-independent interaction, it is possible that more exotic interactions could result in a DM signal at higher energies. This range is also of interest since a significant component of CoGeNT's modulating signal is at higher energies~\cite{Fox:2011px}. The results from this analysis are shown in the lower panels of figure~\ref{fig:likes_energy}. As the three splines are similar at between 1.5 and~$3.0~\keVee$ (lower left panel) the resulting bulk-only spectra in the lower central panel are similar. Furthermore, these spectra are in good agreement with the background estimate (in black) so it is not surprising that the likelihoods do not peak at non-zero values of the cross-section (lower right panel). The reason why the likelihoods for the three splines are different in the lower right panel is because the spline prior is different in each case (the bulk likelihood is similar), reflecting the fact that spline 1 and spine 3 give progressively worse fits to the bulk fraction at low energy. The marginalised result is again similar to previous results; after marginalising over all spines we obtain a p-value of $p = 0.93$ and a Bayes factor of $\mathrm{ln} \, \mathcal{B} = -0.80$, consistent with other values (cf.~table~\ref{tab:pvals}).

\section{Marginalisation procedure and other technical details}
\label{sec:mar_details}

This section provides details of our Bayesian method to marginalise over the bulk fraction $\mathcal{R}(E)$. We start with equation~\eqref{eqn:mar1} for the marginalised posterior. This is calculated by comparing CoGeNT data to the bulk background plus a potential DM recoil for every possible form of the bulk fraction~$\mathcal{R}(E)$. The marginalisation is then achieved by summing up all of these likelihoods, multiplied by a prior for $m$, $\sigma$ and $\mathcal{R}(E)$. The latter prior provides a weight for each spline in the marginalisation integral.
 
We now describe how we incorporated the information from the values of $\mathcal{R}$ into our marginalisation procedure. We model~$\mathcal{R}(E)$ as a cubic spline, parameterised by a set of knots with positions $x_i = (E_i,\mathcal{R}_i)$ on the $E$-$\mathcal{R}$ axis, as shown for example in figure~\ref{fig:bulk_frac}. Parameterising~$\mathcal{R}(E)$ with cubic splines is a good choice in the absence of any theoretically motivated function. The quality of the fit of the spline to data for $\mathcal{R}$ is measured by the posterior
\begin{equation}
\mathcal{P}(\mathcal{R}(E) | d_{\mathcal{R}}) = \frac{ \mathcal{P}(d_{\mathcal{R}} | \mathcal{R}(E)) \prod_i \mathcal{P}(x_i)  }{\mathcal{P}(d_{\mathcal{R}})},
\label{eqn:risetime_post}
\end{equation}
where the priors for each of the knot positions $\mathcal{P}(x_i)$ are assumed to be constant. We use this directly for our prior on $\mathcal{R}(E)$, i.e.~we set $\mathcal{P}(\mathcal{R}) \equiv \mathcal{P}(\mathcal{R}(E) | d_{\mathcal{R}})$. Inserting this into equation~\eqref{eqn:mar1} gives
\begin{equation}
\mathcal{P}(m,\sigma | d) \mathcal{P}(d)  = \int \mathcal{P} (d_E | m,\sigma,\mathcal{R}(E))\mathcal{P}(d_{\mathcal{R}} | \mathcal{R}(E)) \mathcal{P}(m,\sigma)  \prod_i \mathcal{P}(x_i) \, \mathrm{d} x_i,
\label{eqn:mar2}
\end{equation}
where the integral over $\mathcal{R}$ has been converted into an integral over the positions of the knots. So, for example, $i$ runs between one and four for a 4-knot spline and the integral over $\mathcal{R}$ becomes a four-dimensional integral.

In practice, this integral is evaluated as a discrete sum. Setting all of the priors to be constant within some finite space reduces equation~\eqref{eqn:mar2} to the discrete sum in equation~\eqref{eqn:mar_discrete}. This means that we prepare a set of splines which we wish to marginalise over, defined by their knot positions, and then calculate $\mathcal{P}(d_{\mathcal{R}} | \mathcal{R}(E))$ and $\mathcal{P} (d_E | m,\sigma,\mathcal{R}(E))$ using each spline one-by-one. We then sum the products of these likelihoods and the constant prior $\mathcal{P}(m,\sigma)$ to obtain the marginalised posterior. It is important that the range of splines (or other suitable functions) scanned over in this sum is large enough to cover the range of possible choices, otherwise the result will be biased. 

As discussed in section~\ref{sec:1129}, we allow the amplitude of the knots together with the energy of the two lowest energy-knots to float. When a log-normal distribution is used to model the bulk events and $\tau_{\rm{cut}}=5~\mu$s, the $E$-axis position is varied between~$0.5$ and~$1.2$~keV$_{\mathrm{ee}}$ for the knot at lowest energy, and between~$1.2$ and~$2.0$~keV$_{\mathrm{ee}}$ for the knot at second lowest energy. In both cases, the positions are varied in units of 0.1~keV$_{\mathrm{ee}}$. The $\mathcal{R}$ value of the two lowest energy knots are varied between $0$ and $1$, within $50$ evenly-spaced bins. A similar procedure is followed for the Pareto distribution when $\tau_{\rm{cut}}=5~\mu$s. The difference is that the lowest energy knot is varied between~$0.5$ and~$1.0$~keV$_{\mathrm{ee}}$ while the next knot is varied between~$1.0$ and~$1.6$~keV$_{\mathrm{ee}}$.

In the cases where we vary the rise-time cut, $\mathcal{R}(E)$ is replaced with the corrected bulk fraction $\mathcal{R}_c(E)$, which incorporates an efficiency factor (see equation~\eqref{eqn:R_corr_1}). Since $\mathcal{R}_c$ can take values greater than unity, we scan between $0.5$ and $4.0$ in $50$ separate bins, while also allowing the positions of the two lowest-energy knots to vary on the $E$-axis, between $0.5$ and $1.1$~keV$_{\mathrm{ee}}$ for the lowest knot and between~$1.1$ and~$1.8$~keV$_{\mathrm{ee}}$ for the second-lowest. The other knots do not vary with $E$.

\section{Bayes factors and p-values}
\label{sec:numbers}

Our analysis includes both Bayesian and frequentist statistical tests. For the Bayesian case, we can determine to what extent the data prefer a DM+background fit over the background-only scenario using the Bayes factor~\cite{Arina:2012dr,Arina:2013jma}. This is defined as
\begin{equation}
\mathcal{B} = \frac{\int \mathrm{d} m \, \mathrm{d} \sigma \,  \mathrm{d} \mathcal{R} \,  \mathcal{P} (d_E | m,\sigma,\mathcal{R}(E)) \mathcal{P}(\mathcal{R}) \mathcal{P}(m,\sigma)}{\int \mathrm{d} \mathcal{R} \,  \mathcal{P} (d_E | \sigma = 0,\mathcal{R}(E)) \mathcal{P}(\mathcal{R})},
\label{eqn:bayes_factor}
\end{equation}
where all terms are as defined in section~\ref{sec:mar}, and $\mathcal{P} (d_E | \sigma = 0,\mathcal{R}(E))$ is the likelihood function in the case of zero DM signal (i.e.~the background-only scenario). We have marginalised over the bulk-fraction~$\mathcal{R}(E)$ using the prior from the spline fit to the~$\mathcal{R}$ data points. In all cases, we define our prior $\mathcal{P}(m,\sigma)$ to be linearly flat in mass between 2 GeV and 20 GeV (we focus on low-mass DM) and logarithmically flat in $\sigma$ between $10^{-46}$ cm$^2$ and $10^{-34}$ cm$^2$ (the priors are zero outside these ranges). We have tested our conclusions with different priors, and our results do not change significantly.

If $\mathcal{B} > 1$ then the fit to the data is improved when adding a DM recoil signal to the bulk background spectrum (after marginalising over $\mathcal{R}(E)$), while $\mathcal{B} < 1$ implies that the data disfavours a DM recoil signal  \cite{Arina:2012dr}. It is generally more convenient to work with the logarithm of $\mathcal{B}$, in which case $-1 < \mathrm{ln}\,\mathcal{B} < 0$ implies weak/inconclusive preference towards the background-only scenario, and $-3 < \mathrm{ln}\,\mathcal{B} < -1$ implies substantial preference. For positive values of $\mathrm{ln}\,\mathcal{B}$, the DM recoil scenario is preferred over background.

Our discussion can also be placed in a frequentist context. We use a profile likelihood method to profile out the bulk fraction~$\mathcal{R}(E)$, analogous to marginalisation in the Bayesian method. After this, we use a likelihood ratio test to compare the DM+background with the background-only scenario, and define p-values from this quantity. In full, this profiled likelihood ratio is
\begin{equation}
\lambda = -2 \ln \, \left[ \frac{\mathrm{max} \, \mathcal{P} (d_E | m,\sigma,\mathcal{R})  \mathcal{P}(d_{\mathcal{R}} | \mathcal{R})}{\mathrm{max} \, \mathcal{P} (d_E | \sigma = 0,\mathcal{R})  \mathcal{P}(d_{\mathcal{R}} | \mathcal{R})} \right],
\end{equation}
where `max' refers to the largest value of the numerator for all values of $\mathcal{R}$, $m$ and $\sigma$, and the maximum value of  the background-only likelihood $\mathcal{P} (d_E | \sigma = 0,\mathcal{R})  \mathcal{P}(d_{\mathcal{R}} | \mathcal{R})$ for all splines.

We assume that $\lambda$ follows a $\chi^2$ distribution with two degrees of freedom (mass and cross-section), which allows us to convert $\lambda$ to a p-value. Larger values of $\lambda$ imply smaller values of $p$, with for example $p = 0.32$ indicating a $1\sigma$ preference for DM over the bulk background only, $p = 0.05$ indicating a $2\sigma$ preference and so on.

\bibliography{cogent_bayes}
\bibliographystyle{JHEP}

\end{document}